\documentclass[journal, final]{IEEEtran}

\usepackage[lined,linesnumbered,ruled]{algorithm2e}
\makeatletter
\newcommand{\removelatexerror}{\let\@latex@error\@gobble}
\makeatother
\SetAlFnt{\footnotesize\sffamily}

\usepackage{cite}
\usepackage{subfigure}
\usepackage{graphicx}
\usepackage{xcolor}
\usepackage{multirow}
\usepackage{xcolor,colortbl}

\usepackage{amsmath, amsthm, amssymb}

\usepackage[none]{hyphenat}
\begin{document}
\makeatletter
\@addtoreset{footnote}{page}
\renewcommand{\thefootnote}{\ifcase\value{footnote}\or 1 \or 2 \or 3 \or 4 \or * \fi}
\makeatother
\title{Elastic-TCP: Flexible Congestion Control Algorithm to Adapt for High-BDP Networks.}

\author{Mohamed A. Alrshah$^{1,2*}$ (Senior Member, IEEE), Mohamed A. Al-Moqri$^3$, \\ Mohamed Othman$^{1,4}$ (Senior Member, IEEE)
}

\markboth{IEEE Systems Journal, 2019}%
{Alrshah \MakeLowercase{\textit{et al.}}: Elastic-TCP: Flexible Congestion Control Algorithm to Adapt for High-BDP Networks.}
\maketitle

\footnotetext[1]{Department of Communication Technologies \& Networks, Universiti Putra Malaysia (UPM), 43400, Serdang, Malaysia.}
\footnotetext[2]{Al Asmarya Islamic University, Zliten, Libya}
\footnotetext[3]{Azal University for Human Development, 60th Street, Sana'a, Yemen.}
\footnotetext[4]{Researcher at INSPEM Computational and Mathematical Lab, UPM.}
\footnotetext{$^*$Corresponding authors: Mohamed Alrshah, mohamed.asnd@gmail.com}

\begin{abstract}
In the last decade, the demand for Internet applications has been increased, which increases the number of data centers across the world. These data centers are usually connected to each other using long-distance and high-speed networks. As known, the Transmission Control Protocol (TCP) is the predominant protocol used to provide such connectivity among these data centers. Unfortunately, the huge Bandwidth-Delay Product (BDP) of these networks hinders TCP from achieving full bandwidth utilization. In order to increase TCP flexibility to adapt for high-BDP networks, we propose a new delay-based and RTT-independent Congestion Control Algorithm (CCA), namely Elastic-TCP. It mainly contributes the novel Window-correlated Weighting Function (WWF) to increase TCP bandwidth utilization over high-BDP networks. Extensive simulation and testbed experiments have been carried out to evaluate the proposed Elastic-TCP by comparing its performance to the commonly used TCPs developed by Microsoft, Linux, and Google. The results show that the proposed Elastic-TCP achieves higher average throughput than the other TCPs, while it maintains the sharing fairness and the loss ratio. Moreover, it is worth noting that the new Elastic-TCP presents lower sensitivity to the variation of buffer size and packet error rate than the other TCPs, which grants high efficiency and stability.
\end{abstract}


\IEEEpeerreviewmaketitle

\section{Introduction}\label{Intro}
Recently, the demand for Internet applications has been increased, which increases the number of data centers across the world. In order to improve the connectivity between these data centers, high-speed and long-distance networks are widely used across many countries and continents. As known, Transmission Control Protocol (TCP) is the main protocol used to provide an efficient connectivity among these data centers. Unfortunately, the huge Bandwidth-Delay Product (BDP) of these high-speed and long-distance networks hampers TCP from fully utilizing bandwidth, which is considered as a waste of very expensive and important network resources \cite{Afanasyev2010, Scharf2011, xu2011, Callegari2012b, Callegari2014, wang2016tcp, cardwell2017, rhee2018cubic}.

Indeed, high-BDP networks are not a typical environment for which most TCP Congestion Control Algorithms (CCAs) are designed. Specifically, this environment causes two major unavoidable problems that negatively affect the general performance of TCP. The first problem is the long Round Trip Time (RTT) caused by long-distances of network links and by applying big buffer regimes. The second problem is the need for expanding the congestion window (\textit{cwnd}) to a big number of packets in order to utilize the available bandwidth due to the high-BDP of these networks. In the congestion avoidance stage, TCP requires around an RTT to increase its \textit{cwnd} by one and because the RTT in such networks is long, thus, the increase of \textit{cwnd} becomes severely slow \cite{xu2011, alrshah2014, alrshah2017, kanagarathinam2018d}.

As a result of the aforementioned two problems, TCP spends a long period of time to grasp the maximum capacity of high-BDP links, which under-utilizes the network bandwidth. Moreover, after reaching the maximum bandwidth limit, congestion losses (periodically happen) cause an acute \textit{cwnd} degradation. In turn, TCP requires additional time to reach the maximum \textit{cwnd} again, which increases its sensitivity to packet loss. In the recent years, many TCP CCAs have been suggested to solve the aforementioned problems. Although these TCP CCAs have made many improvements, they are still incapable to efficiently utilize the available bandwidths of such high-BDP links and even they present a very high sensitivity to packet loss \cite{Afanasyev2010, Scharf2011, xu2011, Callegari2012b, Callegari2014, wang2016tcp, cardwell2017, rhee2018cubic, alrshah2014, acharya2012, alrshah2009}. 

This paper proposes a new delay-based and RTT-independent TCP CCA, namely Elastic-TCP, which mainly contributes the novel Window-correlated Weighting Function (WWF) in order to augment the bandwidth utilization over high-BDP networks. The WWF improves the ability of Elastic-TCP to deal with big buffers, long delays and high-BDP networks. Extensive simulation and testbed experiments have been carried out to evaluate the proposed Elastic-TCP compared to C-TCP, Cubic, Agile-SD, and TCP-BBR.

The remainder of this article is coordinated as follows: the related works are presented in Section \ref{RW} while the proposed Elastic-TCP is exhibited in Section \ref{Algo}. Sections \ref{PE} and \ref{PET} show the performance evaluation based on simulation and testbed, respectively. Finally, Section \ref{disc} presents the summary and discussion of results, and Section \ref{Conc} concludes the work and points out the future directions.

\section{Related Works}
\label{RW}
In the recent years, many CCAs have been developed to solve network congestion problems and also to enhance the overall performance of TCP, especially in high-BDP networks. Table \ref{history} shows the historical development of TCP CCAs designed for high-speed networks. 

\begin{table} [t]
	\caption {TCP variants designed for H-BDP networks and their implementations in popular operating systems \cite{Afanasyev2010,alrshah2014}.}
	\begin{center}{\scriptsize
		\begin{tabular}{p{0.3cm} p{1.55cm}p{2.6cm}p{0.7cm}p{0.65cm}p{0.5cm}}
			\hline
			\rowcolor[HTML]{FFFFFF} Date	& TCP CCA						& Based on 				& Windows    	& Linux	      	& Solaris 	\\ \hline
			\rowcolor[HTML]{EFEFEF} 1999	& NewReno \cite{floyd1999}		& Reno      			& *NI    		& $>$2.1.36 	& 7.0  		\\ 
			\rowcolor[HTML]{FFFFFF} 2003	& HS-TCP \cite{Floyd2003}      	& NewReno   			& *NI    		& $>$2.6.13 	& *NI   	\\ 
			\rowcolor[HTML]{EFEFEF} 2003	& S-TCP \cite{Kelly2003}  	    & NewReno   			& *NI    		& $>$2.6.13 	& *NI   	\\ 
			\rowcolor[HTML]{FFFFFF} 2003	& Fast \cite{jin2003}     		& Vegas     			& *NI    		& *NI			& *NI   	\\ 
			\rowcolor[HTML]{EFEFEF} 2004	& H-TCP \cite{Leith2004}    	& NewReno   			& *NI    		& $>$2.6.13		& *NI   	\\ 
			\rowcolor[HTML]{FFFFFF} 2004	& Hybla \cite{Caini2004}     	& NewReno   			& *NI    		& $>$2.6.13 	& *NI   	\\ 
			\rowcolor[HTML]{EFEFEF} 2004	& BIC-TCP \cite{xu2004}			& HS-TCP    			& *NI    		& $>$2.6.12 	& *NI   	\\ 
			\rowcolor[HTML]{FFFFFF} 2005	& AFRICA \cite{King2005}    	& HS-TCP, Vegas			& *NI    		& *NI   		& *NI   	\\ 
			\rowcolor[HTML]{EFEFEF} 2005	& NewVegas \cite{JoelSing2005} 	& Vegas					& *NI			& *NI			& *NI		\\ 
			\rowcolor[HTML]{FFFFFF} 2005	& AReno \cite{shimonishi2006} 	& Westwood,	Vegas		& *NI			& *NI			& *NI		\\ 
			\rowcolor[HTML]{EFEFEF} 2006	& C-TCP \cite{Tan2006}       	& NewReno,HS-TCP,Vegas 	& V,7,8,10  	& $>$2.6.14 	& *NI 	  	\\ 
			\rowcolor[HTML]{FFFFFF} 2006	& illinois \cite{Liu2008}  		& NewReno, DUAL			& *NI     		& $>$2.6.22 	& *NI	   	\\ 
			\rowcolor[HTML]{EFEFEF} 2007	& Fusion \cite{Kaneko2007}    	& Westwood, Vegas		& *NI    		& *NI	  		& $>$10		\\ 
			\rowcolor[HTML]{FFFFFF} 2007	& YeAH \cite{Baiocchi2007}    	& S-TCP, Vegas			& *NI     		& $>$2.6.22 	& *NI 	  	\\ 
			\rowcolor[HTML]{EFEFEF} 2008	& Cubic \cite{Ha2008}		 	& BIC-TCP, H-TCP		& *NI     		& $>$2.6.16		& *NI		\\ 
			\rowcolor[HTML]{FFFFFF} 2015	& Agile-SD \cite{alrshah2015}   & NewReno		   		& *NI     		& $\geq$4.0    	& *NI   	\\ 
			\rowcolor[HTML]{EFEFEF} 2017	& TCP-BBR \cite{cardwell2017} 	& Vegas  				& *NI     		& $\geq$4.9   	& *NI   	\\ \hline
		\end{tabular}}
		\begin{flushleft}\vspace{-0.1cm}\hspace{0.2cm}\scriptsize *NI = Not Implemented\end{flushleft}
		\label{history}
	\end{center}\vspace{-1cm}
\end{table}


In high-BDP networks, loss-based CCAs are very sensitive to packet loss, and delay-based CCAs are highly sensitive to RTT changes, while RTT-dependent CCAs are suffering from severe throughput degradation and low fairness \cite{Afanasyev2010, alrshah2014, alrshah2015}. RTT-dependent CCAs increase their \textit{cwnd}, at congestion avoidance stage, by one every RTT. Thus, if the RTT is small the increase will be fairly fast otherwise it will be unacceptably slow. In fact, RTT-dependency causes unfair share among competing flows that have different RTT lengths, in which the shorter the RTT the higher the aggressiveness and vice versa. RTT-dependency also increases the sensitivity to packet loss and negatively influences the overall performance of TCP \cite{Afanasyev2010, Callegari2012b, alrshah2014, alrshah2015}. For these reasons, RTT-independent CCAs are highly recommended for high-BDP networks. RTT-independency allows TCP to increase its \textit{cwnd} based on the changes of underlaying network instead of RTT magnitude, which significantly improves throughput.

In 2006, C-TCP \cite{Tan2006} proposed a new hybrid CCA, which improved the performance of TCP to some extent. However, it inherits the RTT estimation problem from TCP Vegas \cite{brak1995}, which increases its sensitivity to RTT changes and negatively affects the fairness. Moreover, C-TCP is also an RTT-dependent CCA, which makes the growth of its \textit{cwnd} very slow, notably over high-BDP networks. Despite all, C-TCP has been set as the default TCP for MS Windows since its first implementation in Windows Vista, which makes it one of the most widely used TCP in the world~\cite{Afanasyev2010},~\cite{alrshah2014}.

In 2008, Cubic \cite{Ha2008} became the default TCP of the afterward versions of Linux kernel. It improved the scalability over high-BDP networks by increasing its \textit{cwnd} in the congestion avoidance stage using $cubic$ root of the elapsed time since last loss. However, it becomes a time-consuming protocol since it is an RTT-dependent TCP, which results in an underutilization of bandwidth over high-BDP networks \cite{Afanasyev2010, alrshah2014, alrshah2015}.

In 2017, Agile-SD \cite{alrshah2017} was proposed to reduce the sensitivity to packet loss and to grant the ability to deal with small buffers over high-speed networks. Agile-SD was designed for short-distance networks, where the delay-based approach is not functioning due to the triviality of RTT variation in such networks. Despite that Agile-SD significantly improved the performance over short-distance networks, it still has a limited performance over high-BDP networks.

In 2018, TCP-BBR \cite{cardwell2017} was proposed by a research group at Google as a model-based CCA. It estimates the bottleneck, bandwidth, and RTT in order to improve the link utilization while keeping the bottleneck queue un-congested. Despite the implementation of TCP-BBR in Google and YouTube Infrastructure, it is still suffering from maintaining un-congested queue at the expense of bandwidth utilization. Specifically, if a TCP-BBR flow concurrently shares a bottleneck with another Cubic flow, the latter will aggressively fill up the queue while the former will trigger its draining function to empty that queue. Consequently, TCP-BBR flows will get smaller share compared to Cubic flows. On the other hand, TCP-BBR will not properly function for short-term flows, such as request/response flows, since TCP-BBR needs many cycles to estimate its parameters. Moreover, TCP-BBR presents a very high level of code complexity compared to other algorithms, as shown in Table~\ref{code}.

\begin{table}[h]
	\vspace{-0.4cm}
	\caption{Number of code lines for the studied algorithms in terms.}\vspace{-0.6cm}
	\begin{center}
		\begin{tabular}{c|ccccc}\hline 
			\rowcolor[HTML]{FFFFFF} CCA 		& TCP-BBR 	& Cubic & C-TCP & Elastic-TCP & Agile-SD 	 \\ \hline 
			\rowcolor[HTML]{EFEFEF} Code lines  & 553 		& 342 	& 219 	& 149 		  & 115 		 \\	\hline 
		\end{tabular} \label{code}
	\end{center}\vspace{-0.4cm}
\end{table}

At the congestion avoidance stage, most TCP CCAs increase their \textit{cwnd} by $Inc$, which varies from CCA to another. This $Inc$ is calculated based on different parameters, such as predefined constants and time, depends on the applied CCA. If the \textit{cwnd} is small (short-distance), the increase will be reasonably fast and even aggressive sometimes. However, if the \textit{cwnd} is large (long-distance), the increase will be severely slow. The main cause of this problem is that TCP does not correlate the value of $Inc$ to the magnitude of the \textit{cwnd} itself.

For example, Reno and NewReno calculate their $Inc$ as $\frac{\alpha}{cwnd}$, where $\alpha$ is a predefined constant usually equal~to~1. In C-TCP, $Inc$ is calculated as the sum of $cwnd_{reno}$ and $cwnd_{fast}$, where $cwnd_{reno}$ is the Reno increase as calculated above and $cwnd_{fast}$ is the HS-TCP increase, which is calculated as $cwnd_{fast} - (\zeta . \Delta)$, where $\Delta$ is the Vegas-estimate and $\zeta$ is a predefined constant. As for Agile-SD, $Inc$ is calculated as $\frac{\hspace{0.1cm}\lambda\hspace{0.1cm}}{cwnd}$, where $\lambda$ is dynamically calculated based on the change in \textit{cwnd} and always $\lambda \geq 1$. As for Cubic, $Inc$ is calculated as $C (\Delta - \sqrt[3]{\frac{\beta * cwnd}{C}})^3$, where $C$ is a preset constant and $\beta$ is the multiplicative decrease factor while $\Delta$ indicates the elapsed time since last loss.

Based on the aforementioned $Inc$ calculations, it can be clearly observed that $Inc$ (in Reno, NewReno, C-TCP and Agile-SD) is reversely proportional to \textit{cwnd} with no correlation to the magnitude of that \textit{cwnd}. As for Cubic, the $Inc$ is directly proportional to the \textit{cwnd}, but the greater the magnitude of \textit{cwnd} the smaller the value of $Inc$. Thus, in high-BDP networks, where the magnitude of \textit{cwnd} is very large, the growth of \textit{cwnd} in all studied CCAs is severely slow. In all studied TCPs, the $Inc$ calculations are directly correlated to the predefined constants, which hampers the ability of these TCPs to adapt to both small and large \textit{cwnd} scenarios simultaneously. Consequently, TCP setting which can be appropriate for short-distance networks, is usually improper for high-BDP networks and vice versa.

For better understanding, let us consider an example of NewReno over a low-BDP network link with $1Gbps$ bandwidth, $1ms$ RTT, and $1Kbyte$ packet size. The BDP of this link is approximately 125 packets based on Equation (\ref{eq1})~\cite{Ha2008} below:
\begin{align}
	\label{eq1}
	\textit{BDP(packets)} &= \frac{\textit{Bandwidth(bps) }\times \textit{ RTT(seconds)}}{\textit{Packet Size(bits)}}
\end{align}

Mostly, TCP degrades its \textit{cwnd} to the half of link BDP ($\approx$~62~packets) after congestion occurrence. Then, it starts another epoch using the additive increase (one packet per RTT) to attain the maximum \textit{cwnd} again. Consequently, it consumes 62 RTTs per epoch, which is about 62 milliseconds in this example,  to reach the maximum link BDP. Thus, this behavior gives an acceptable throughput and, in turn, achieves a fair level of bandwidth utilization.

However, if the RTT in the aforementioned example is prolonged to be $100ms$ in order to emulate high-BDP link scenarios, the link BDP will become about 12,500 packets based on Equation (\ref{eq1}). As above-mentioned, TCP decreases its \textit{cwnd} to the half of link BDP \mbox{($\approx$ 6,250 packets)}. In the following epochs, TCP will consume about 6,250 RTTs \mbox{($\approx$~625~seconds)} per epoch to attain the maximum \textit{cwnd}. Thus, this very sluggish behavior degrades the performance and harshly under-utilizes the link bandwidth. Furthermore, when the network bandwidth is increased to $10$, $100Gbps$ or more, such problem will become significantly more severe.

In this work, we propose the Elastic-TCP to enhance the bandwidth utilization over high-BDP networks, in which RTTs are very long, buffers are very large and packet loss are very common. Elastic-TCP is a new delay-based and RTT-independent CCA contributing a novel WWF function that correlates the value of \textit{cwnd} increase to the \textit{cwnd} magnitude. Besides, the gained increase is balanced using the weighting function according to the variation of RTT in order to maintain the fairness. Consequently, this behavior improves the ability of TCP to adapt to different networks with variable \textit{cwnd} magnitudes, which especially improves the bandwidth utilization over high-BDP networks.

\section{Elastic-TCP: The Proposed Algorithm}
\label{Algo}
\begin{figure}[h]
	\centering
	\includegraphics[width=0.9\linewidth]{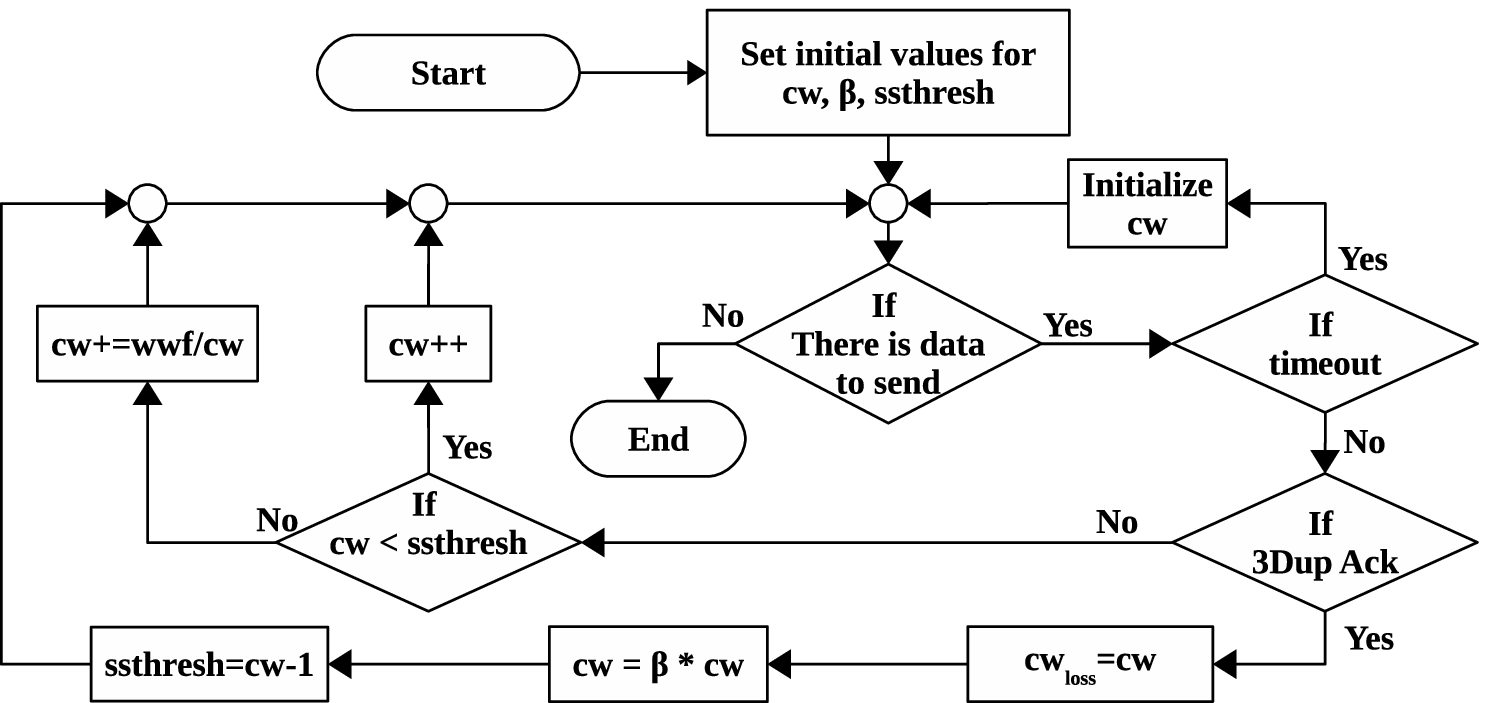}
	\caption{The control flow diagram of Elastic-TCP}
	\label{fig:flowchart}
\end{figure}

Elastic-TCP is a delay-based and RTT-independent CCA designed for high-BDP networks to improve the bandwidth utilization without jeopardizing the fairness. For more details, Figure \ref{fig:flowchart} shows the control flow diagram of the proposed Elastic-TCP and Algorithm \ref{algo01} describes the internal functionality of it while the following subsections provide a deep explanation of its unique mechanism.

\begin{figure}[!h]
	\removelatexerror
	\centering
	\begin{minipage}{7.5cm}
		\SetKwProg{Function}{Function}{ }{end}
		\SetKwProg{Event}{Event}{ do}{end}
		\begin{algorithm}[H]	
		\caption{The pseudocode of Elastic-TCP.}\label{algo01}	
		\textbf{Initialization:}\\
		\hspace{0.5cm}$RTT_{max} \leftarrow 0, RTT_{current} \leftarrow 0,$\\
		\hspace{0.5cm}$RTT_{base} \leftarrow \text{0x7FFFFFFF},$ \hspace{0.5cm}$cwnd \leftarrow 2$\\
		\Event{On ACK Receiption}
		{
			\uIf {Not duplicated ACK}
			{
				\uIf {Slow Start}
				{
					$cwnd \leftarrow cwnd + 1$
				}
				\Else
				{
					$RTT_{current} \leftarrow (now - sendtime)$\\
					
					\If {$RTT_{current} < RTT_{base}$}
					{
						$RTT_{base} \leftarrow RTT_{current}$
					}
					
					\If {$RTT_{current} > RTT_{max}$}
					{
						$RTT_{max} \leftarrow RTT_{current}$
					}
					
					$WWF \leftarrow \sqrt{\frac{RTT_{max}}{RTT_{current}} \times cwnd}$
					
					$cwnd \leftarrow cwnd + \frac{WWF}{cwnd}$
				}
			}
			\Else 
			{
				Apply the multiplicative decrease.
			}
		}
		\end{algorithm}
\end{minipage}\vspace{-0.5cm}
\end{figure}

\subsection{Window-correlated Weighting Function (WWF)}
WWF is the primary contribution of this work. Substantially, WWF aims at improving TCP bandwidth utilization over high-BDP networks without jeopardizing the fairness. Elastic-TCP relies on the variation of RTT to measure the utilization ratio ($UR$), which is calculated and used in a different way other than those ways presented by TCP-Dual, Vegas, and Fast-TCP.

\begin{figure} [htpb]
	\centering
	\includegraphics[width=0.8\linewidth]{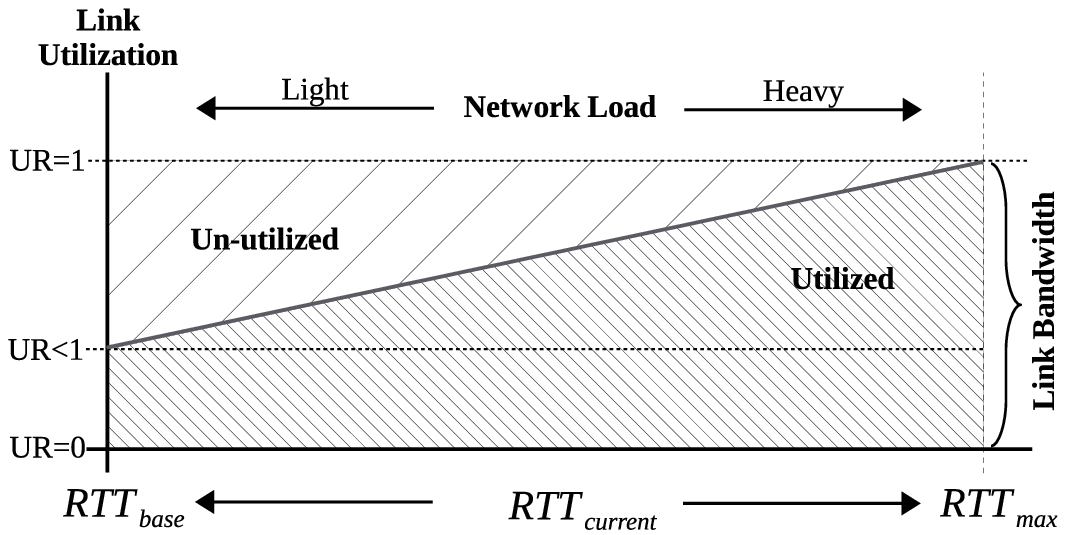}
	\caption{The impact of RTT on UR.}
	\label{fig:ur}
\end{figure}

As known, the variation of RTT can be used to quantify the level of congestion and/or the level of link utilization at the bottleneck \cite{Afanasyev2010, jin2003, brak1995, wang1992}. In this work, we defined the utilization ratio ($UR$) as a percentage of the utilized buffer and BDP, as shown in Figure \ref{fig:ur}. Thus, the proposed Elastic-TCP quantifies the $UR$ at the bottleneck link as in Equation \eqref{eq11010} below: 
\begin{align}
	\label{eq11010}
	UR = \frac{RTT_{current}}{RTT_{max}},
\end{align}
where $RTT_{current}$ is the current RTT obtained from the last ACK, $RTT_{base}$ and $RTT_{max}$ are the minimum and maximum RTT seen over this connection, respectively, where ($RTT_{base} \leq RTT_{current} \leq RTT_{max}$), ($RTT_{base} > 0$), ($RTT_{max} > RTT_{base}$) and ($UR \in [0, 1]$).

\begin{figure} [htpb]
	\centering
	\includegraphics[width=0.75\linewidth]{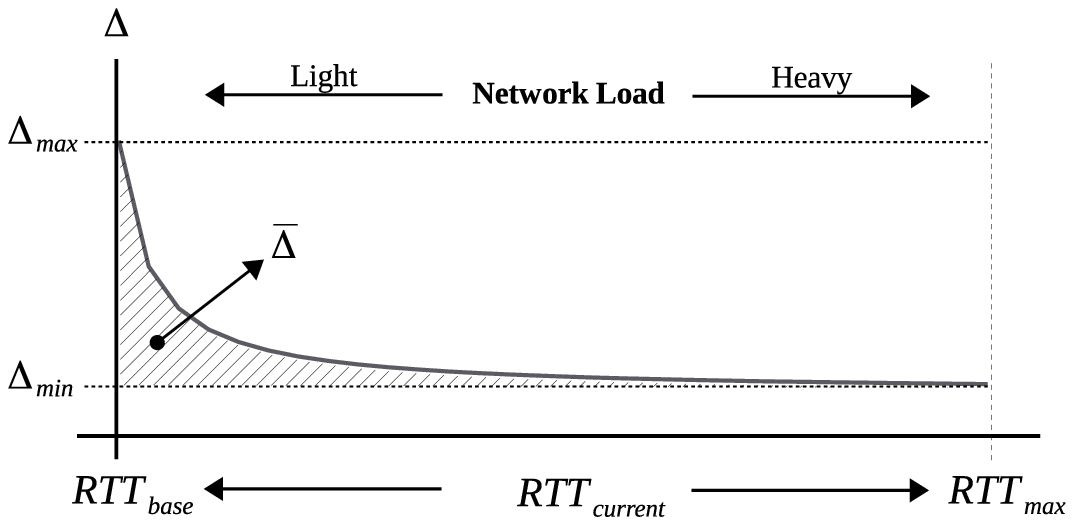} 
	\caption{The impact of RTT on $\Delta$.}
	\label{fig:delta}
\end{figure}

Hence, the underutilization ratio ($\overline{UR}$), which reflects the under-utilized portion of BDP plus the empty buffer size, can be quantified using Equation \eqref{eq110108}: 
\begin{align}
	\label{eq110108}
	\overline{UR} = \frac{RTT_{max} - RTT_{current}}{RTT_{max}} = 1 - UR,
\end{align}
where $UR=1$ only when the bandwidth and buffer at the bottleneck link are fully utilized because the $RTT_{current}$ approaches the maximum delay ($RTT_{max}$) only when the bottleneck link capacity and buffer are about to be full, which results in ($\overline{UR}=0$), as shown in Figure \ref{fig:ur}. Then, the $UR$ is used to calculate the weighting function ($\Delta$), as $\Delta = \frac{1}{UR}$, where $\Delta = 1$ only when $UR = 1$, and $\Delta > 1$ otherwise. Hence, the under-utilized portion of bandwidth at the bottleneck $(\bar{\Delta})$, as shown in Figure \ref{fig:delta}, can be calculated as $\bar{\Delta} = \Delta - 1$.

It is very clear that $\Delta$ is inversely proportional to $RTT_{current}$, which exhibits a semi-hyperbolic curve, as shown in Figure \ref{fig:delta}. In other words, $\Delta$ is enlarged, up to the maximum possible value ($\Delta_{max}$), when the $RTT_{current}$ moves towards the $RTT_{base}$, which indicates to light traffic loaded network, as shown in Equation (\ref{eq3}) below:
\begin{align}
	\Delta_{max} &= \lim_{RTT_{current} \to RTT_{base}} \frac{RTT_{max}}{RTT_{current}} = \frac{RTT_{max}}{RTT_{base}} \label{eq3}
\end{align}

Contrarily, $\Delta$ is shrunk, up to the minimum possible value ($\Delta_{min}$), if the $RTT_{current}$ moves towards the $RTT_{max}$, which indicates to heavy traffic loaded network, as shown in \mbox{ Equation (\ref{eq4})} below:
\begin{align}
	\Delta_{min} &= \lim_{RTT_{current} \to RTT_{max}} \frac{RTT_{max}}{RTT_{current}} = 1 \label{eq4}
\end{align}

The main purpose of $\Delta$ is to estimate the maximum possible \textit{cwnd} $(cwnd_{est})$ for the underlying network, which is calculated as $cwnd_{est} = \Delta \times cwnd.$ Since $\Delta = 1 + \bar{\Delta}$, thus $cwnd_{est} = cwnd + (\bar{\Delta} \times cwnd)$, which always guarantees that ($cwnd_{est} \geq cwnd$). In order to increase the adaptability of Elastic-TCP to deal with different scenarios of diverse \textit{cwnd} magnitudes, the value of the increase in \textit{cwnd} should be correlated to the magnitude of $cwnd_{est}$.
	
\begin{figure} [h!]
	\centering
	\includegraphics[width=0.6\linewidth]{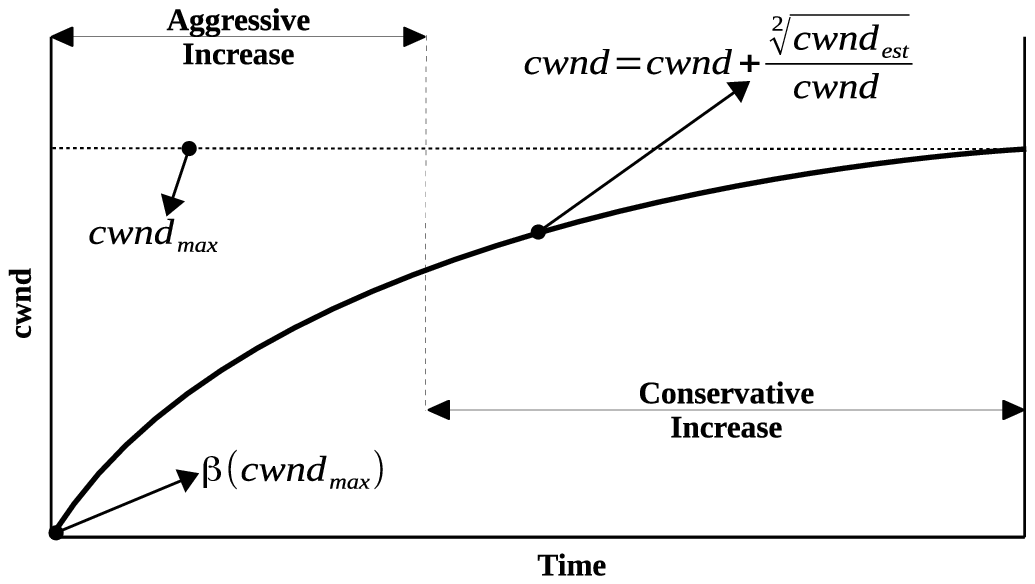} 
	\caption{The window growth function of Elastic-TCP using the square root.}
	\label{fig:sqrt}
\end{figure}

The correlation function should create a convex-up curve to reduce the under-utilized area above the curve, where the more the convexity the best the utilization. However, increasing the convexity more than necessary will lead to severe data loss. Further, the function should grow aggressively when the current \textit{cwnd} is close to the multiplicative decrease point $(\beta*cwnd_{max})$ and should grow conservatively when the current \textit{cwnd} is approaching the maximum bottleneck capacity or the maximum \textit{cwnd} $(cwnd_{max})$, as shown in Figure \ref{fig:sqrt}. Furthermore, the needed function must be a low-complexity function since it will be implemented in the core space of the Linux kernel, which does not provide any high-level user-defined function. For these reasons, we have been searching for a new window growth function that is able to satisfy the above-mentioned constraints. We tested some functions, where we found that the square-root function is able to fulfill the requirements. Thus, we implemented Newton-Raphson iteration method to calculate the square root of $cwnd_{est}$ as \mbox{$WWF = \sqrt{cwnd_{est}}$.}

Finally, the resulted value of WWF is used, in the stage of congestion avoidance, to increase the \textit{cwnd}, as shown in Equation \eqref{eq5} below:
\begin{align}
	cwnd &= cwnd + \dfrac{WWF}{cwnd} \label{eq5}
\end{align}

\begin{figure*} [htpb]
	\centering
	\begin{center}
		\subfigure [Elastic-TCP vs. NewReno]
		{
			\includegraphics[width=0.31\linewidth]{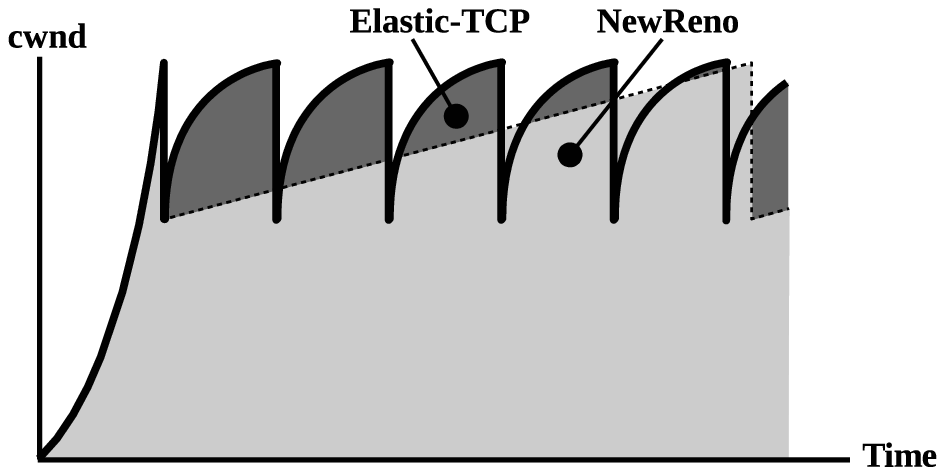}
			\label{fig:newreno}
		}
		\subfigure [Elastic-TCP vs. Cubic]
		{
			\includegraphics[width=0.31\linewidth]{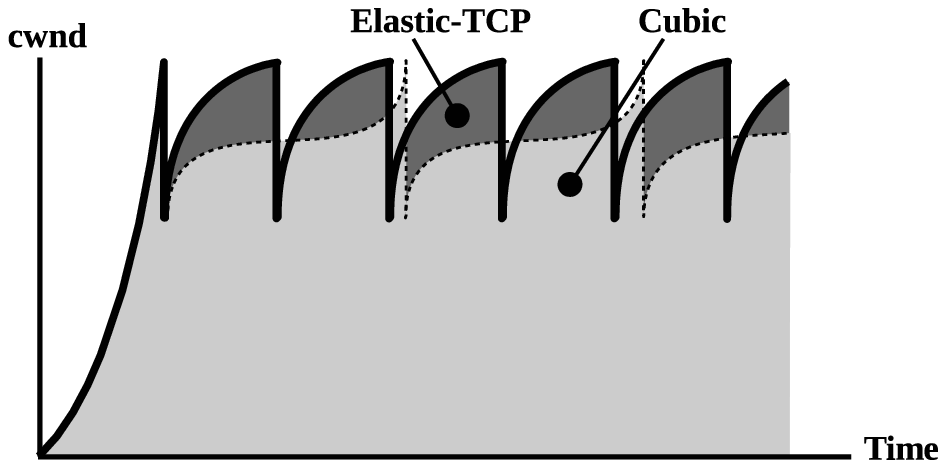}
			\label{fig:cubic}
		}
		\subfigure [Elastic-TCP vs. C-TCP]
		{
			\includegraphics[width=0.31\linewidth]{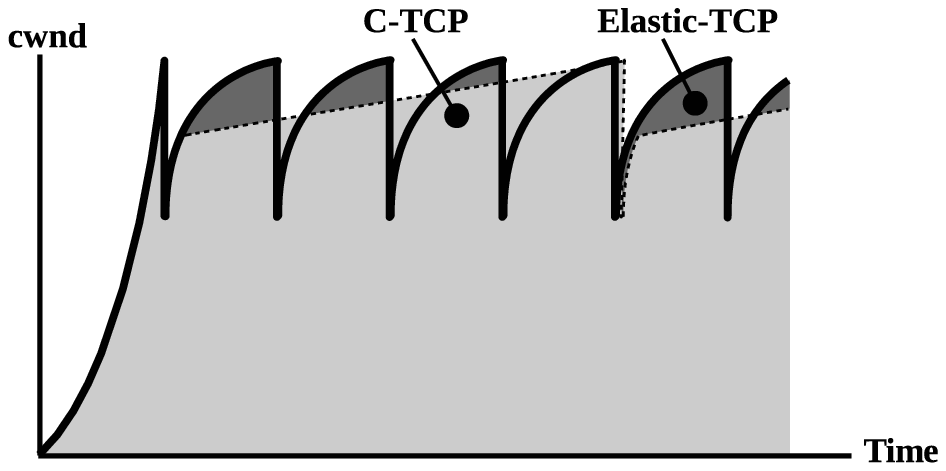}
			\label{fig:ctcp}
		}
	\end{center}
	\caption{The epoch time of Elastic-TCP compared to NewReno, Cubic and C-TCP.}
	\label{fig:cwnd-evo}
\end{figure*}

By this behavior, the novel Elastic-TCP increases its ability to probe the status of the underlying network, as shown in figures \ref{fig:ur} and \ref{fig:delta}. Also, this behavior results in a convex-up curve of increase, in the congestion avoidance stage, which cuts down the epoch time in order to diminish the area of under-utilized bandwidth, as shown in Figure \ref{fig:cwnd-evo}. Specifically, this behavior makes the fast-recovery stage of the Elastic-TCP much faster compared to (1) NewReno, as in Figure \ref{fig:newreno}, (2) Cubic, as in Figure \ref{fig:cubic}, and (3) C-TCP, as in Figure \ref{fig:ctcp}. Besides, this behavior grants low sensitivity to packet losses. Hence, it is very clear that the Elastic-TCP guarantees higher bandwidth utilization and lower sensitivity to packet loss degradation than the existing CCAs while it maintains the fairness.

\subsection{The Elastic-TCP Overall Behavior}
Elastic-TCP starts exponentially as it uses the standard slow start. Then, after detecting the first loss, either by receiving 3-duplicate acknowledgments or by an expiration of the timeout counter, it reduces its \textit{cwnd} by the multiplicative decrease factor $(\beta)$, then it enters the stage of congestion avoidance. In this stage, the Elastic-TCP increases its \textit{cwnd} by $\frac{WWF}{cwnd}$, as shown in Equation \eqref{eq5}, to produce short epochs with convex-up curves of increase. If a packet loss occurs in this stage, the Elastic-TCP reduces its \textit{cwnd} using the multiplicative decrease factor $(\beta)$ to start another epoch of the same stage. 

As shown in figures \ref{fig:newreno}, \ref{fig:cubic} and \ref{fig:ctcp}, this behavior helps the Elastic-TCP to increase its \textit{cwnd} faster than the examined TCP CCAs, which obviously improves the bandwidth utilization. That is to say, the faster the \textit{cwnd} growth the higher the bandwidth utilization and vice versa. However, the most important issue is to which limit \textit{cwnd} has to be increased in order to prevent the problem of over injecting data into the network. Fortunately, the new Elastic-TCP has the ability to improve the bandwidth utilization while keeping data loss as low as in NewReno.

\section{Performance Evaluation of Elastic-TCP using Simulation}
\label{PE}

This work aims at developing a new TCP CCA, namely Elastic-TCP, that improves the bandwidth utilization of high-BDP networks, without jeopardizing the fairness among competing TCP flows. For the purpose of performance evaluation, NS-2 network simulator is used. As well-known, NS-2 provides two ways of TCP implementation, either as a simulation-based module or as a Linux-based module. In this work, we implement the Elastic-TCP into NS-2 as a Linux-based module, which is ready for implementation into Linux kernel.

\subsection{Simulation Setup}
\begin{figure} [htpb]
	\begin{center}
		\subfigure[Congestion-free network topology.] 
		{			
			\includegraphics[width=0.65\linewidth, height=0.7cm]{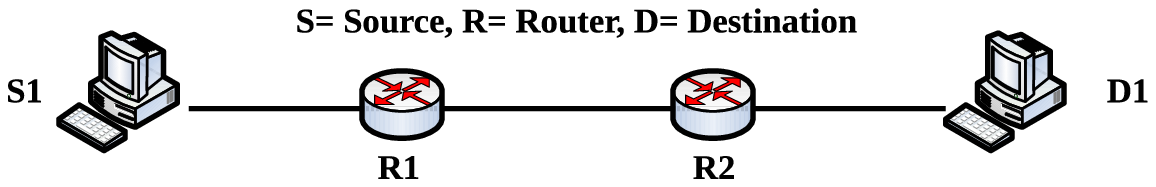}
			\label{fig:topology-ideal}
		}
		\subfigure[Dumbbell Network topology with congested bottleneck.] 
		{			
			\includegraphics[width=0.65\linewidth, height=2.5cm]{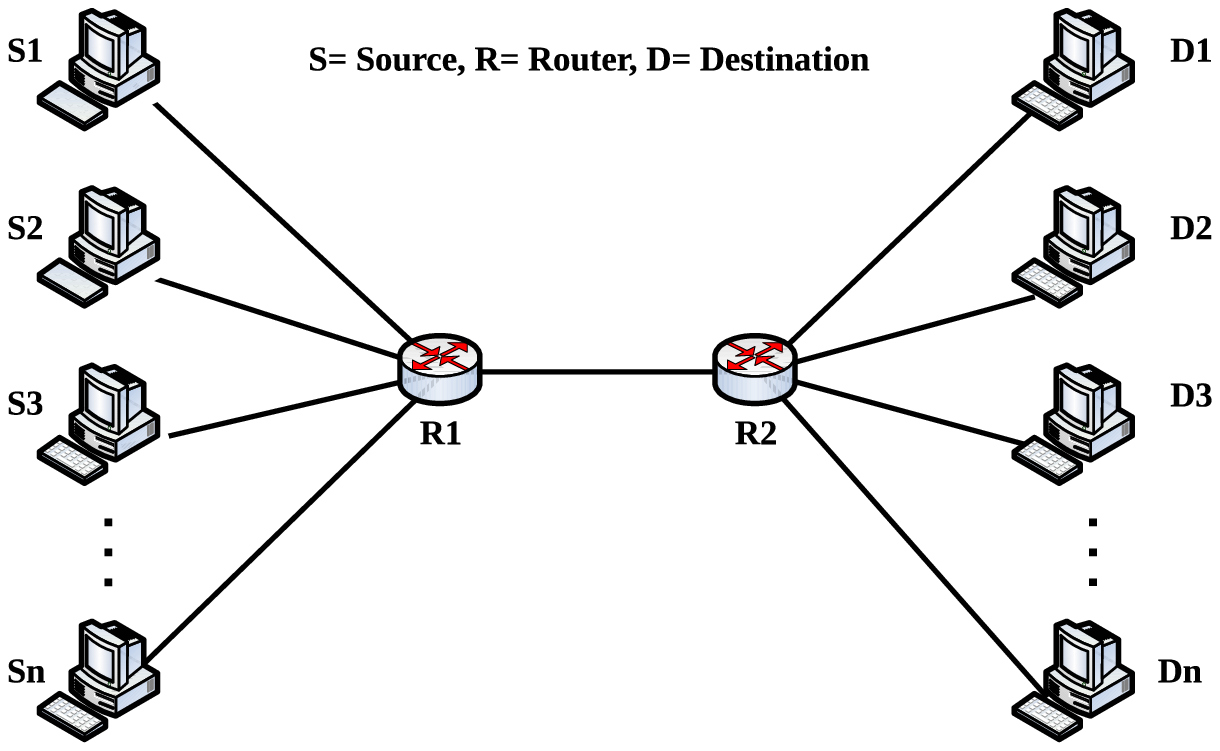}			
			\label{fig:topology}
		}
	\end{center}
	\caption{Network topologies used for performance evaluation in this work.}
\end{figure}

In this paper, NS-2.35 has been used to carry out extensive simulation experiments in order to compare the performance of Elastic-TCP, C-TCP, Cubic, and Agile-SD. The studied algorithms have been examined in three main scenarios:
\begin{enumerate}
	\item Single-flow scenario: this scenario mimics the ideal case of network, which is used to evaluate the performance of TCP over an ideal case of non-congested network, in order to show the maximum achievable bandwidth utilization in the best conditions. This scenario has only one sender and one receiver, the sender starts sending FTP data to the destination from the beginning until the end of simulation.
	
	\item Sequentially established/terminated multiple-flows scenario: it is used to evaluate the performance of TCP over congested bottleneck in order to simulate a real network scenario. This scenario shows the impact of different establishment and termination time of multiple flows on the throughput and on the quality of bandwidth sharing. In this scenario, the flows are established one by one after every 5 seconds starting from time 0 in a manner of point-to-point flows, for example, S1 to D1 at time 0, S2 to D2 at time 5, S3 to D3 at time 10, and so on.
	
	\item Synchronously established/terminated multiple-flows scenario: this scenario shows the impact of synchronized packet loss that occur over all flows on the throughput and on the sharing fairness. In this scenario, all senders start sending FTP data to destinations at the same time (when simulation time $= 0$ sec) and they finish by the end of simulation (when simulation time $= 100$ sec) in a manner of point-to-point flows, for example, S1 sends to D1, S2 sends to D2, and so on.
\end{enumerate}	
In the single flow scenario, the used network topology is as shown in Figure \ref{fig:topology-ideal}, while the topology shown in Figure \ref{fig:topology} is used in multiple-flows scenarios. In the single-bottleneck topology shown in Figure \ref{fig:topology}, $n$ senders compete to send data to $n$ receivers via a shared bottleneck link, where speed and propagation delay are set to $1Gbps$ and $100ms$, respectively. All end-system nodes are linked to bottleneck routers using wired links, where speed and propagation delay are set to $1Gbps$ and $1ms$, respectively \cite{Wang2013}. 

In all scenarios, the performance of the examined TCP CCAs is evaluated with various buffer sizes varying from $50$ to $6400$ packets and Packet Error Rates (PERs) of $10^{-4}$, $10^{-5}$ and $zero$. The buffer size and PER changes only applied to R1 and R2 in order to mimic real bottleneck behavior. As an endeavor to ensure the accuracy of the results, the simulation experiments have been repeated for 30 times for each set of parameters, as shown in Table \ref{params}, then the averages are calculated for each set of parameters.
\begin{table}[htpb]{\footnotesize
	\caption{Simulation parameters setting.}
	\begin{center}
		\begin{tabular}{p{4cm}p{4cm}}
			\hline
			\rowcolor[HTML]{FFFFFF} Parameter							&	Value (s)								\\ \hline
			\rowcolor[HTML]{EFEFEF} TCP CCAs							&	Cubic, C-TCP, Agile-SD, \mbox{Elastic-TCP}		\\ 
			\rowcolor[HTML]{FFFFFF} Link Speed of All Links				&	1Gbps									\\ 
			\rowcolor[HTML]{EFEFEF} PC-to-Router 2-way Delay			&	1 milliseconds							\\ 
			\rowcolor[HTML]{FFFFFF} Bottleneck 2-way Delay				&	100 milliseconds						\\ 
			\rowcolor[HTML]{EFEFEF} Packet Error Rate (PER)				&   $10^{-4}, 10^{-5}, 0$					\\ 
			\rowcolor[HTML]{FFFFFF} Buffer Size at Bottleneck Routers	&	50 to 6400 pckts						\\ 
			\rowcolor[HTML]{EFEFEF} Data Packet Size					&	1 KB									\\ 
			\rowcolor[HTML]{FFFFFF} Management							&	Droptail algorithm						\\ 
			\rowcolor[HTML]{EFEFEF} Flow Type							&	FTP										\\ 
			\rowcolor[HTML]{FFFFFF} Simulation Time						&	100 seconds								\\ 
			\rowcolor[HTML]{EFEFEF} Simulation Runs for Each Scenario	&	30 times								\\ \hline
		\end{tabular}
		\label{params}
	\end{center} }
\end{table}

As well-known, the types of TCP traffic such as HTTP and Telnet are considered as short-lived traffic types, which are not significantly influenced by TCP improvements. In short-lived traffic, tasks are usually accomplished before entering the system steady state. That is why, only FTP traffic is used in this work because it is a long-lived TCP traffic type that represents a great portion of Internet traffic.

Substantially, the aim of these experiments is two-fold: First, to demonstrate the impact of network congestion, variable buffer size, and inconstant PER on the overall performance of examined TCP CCAs. Second, to compare the overall performance of the proposed Elastic-TCP to Cubic, C-TCP and Agile-SD. In all experiments of this work, the simulation time has been set to 100 seconds, which is more than enough for all CCAs to show their behavior in the steady state.

The main goal of this work is to improve the performance of TCP by reducing its sensitivity to packet loss and by increasing its scalability to be able to deal with different networks characteristics. In order to evaluate the performance of TCP at the transport layer, throughput, loss rasio, and sharing fairness index are measured. 

Throughput is the rate of successful data delivery over a network link from sender to receiver. It is usually measured in bits per second (bps) or any unit of its multiples such as Mbps or Gbps. Throughput can be computed as per flow throughput or as system throughput. Say that one TCP flow transmits an amount of data to the receiver side, which received data $(data)$ in bits over a period of time $(time)$ in seconds, thus, the throughput $(Thr)$ of this flow is calculated as $Thr = \frac{data}{time}$. As for the system throughput, suppose that we have a number of flows $(n)$ that send data simultaneously, the system throughput $(SysThr)$ is calculated as below:
\begin{align}
SysThr = \dfrac{\sum_{i=1}^{n} data_i}{time},
\end{align}
where $data_i$ is the data received form the $i^{th}$ flow, and $time$ is the time consumed to receive the $data$ of all flows.

As well-known, data packets can be lost during the data transmission over any type of networks due to many reasons such as congestion, fading, interference. In this work, we count all types of data loss together as one type ($loss$), where this loss is equal to the difference between total data sent ($Sdata$) by a TCP sender and total data received ($Rdata$) by the relative TCP receiver. The loss ratio ($LR$) is calculated as a ratio of data loss to the total data sent ($Sdata$) for all flows ($n$) as calculated below:
\begin{align}
LR = \frac{\sum_{i=1}^{n}SData_i - RData_i}{\sum_{i=1}^{n}SData_i} = \frac{\sum_{i=1}^{n}loss_i}{\sum_{i=1}^{n}SData_i},
\end{align}
where ($Sdata_i$) and ($Rdata_i$) are the total data sent and the total data received for flow ($i$), respectively.

The sharing fairness index is calculated to show whether the competing TCP flows are getting a fair share of the bottleneck link bandwidth. In this work, three types of sharing fairness, namely intra-fairness, RTT-fairness and inter-fairness, are measured using the well-known Jain’s fairness index (JFI) \cite{jain1984}, as shown in Equation \eqref{ch3:fair} below:
\begin{align}
JFI(x_1, x_2, ..., x_n) = \dfrac{(\sum_{i=1}^{n} x_i)^2}{n \cdot \sum_{i=1}^{n} x_i^2} 	\label{ch3:fair},
\end{align}
where ($n$) is the number of flows, and ($ x_i $) denotes the average throughput of the $i^{th}$ flow. Intra-fairness is to measure how fair is the distribution of bottleneck bandwidth among the flows of the same TCP variant, and RTT-fairness is to measure how fair is the distribution of queuing delay among the competing flows originated from the same TCP variant. As for Inter-fairness, it is to measure how fair is the distribution of bottleneck bandwidth among the flows of different TCP variants.

\subsection{Simulation Results and Discussion}
This subsection analytically discusses the behavior shown by Elastic-TCP compared to the other CCAs. Moreover, it presents the performance results in terms of throughput, loss ratio, and fairness in order to exhibit the effect of error rate and buffer size on the overall performance.

\subsubsection{The \textit{cwnd} evolution}
\begin{figure} [h!]
	\begin{center}
		\subfigure[Elastic-TCP]
		{
			\includegraphics[width=0.45\linewidth]{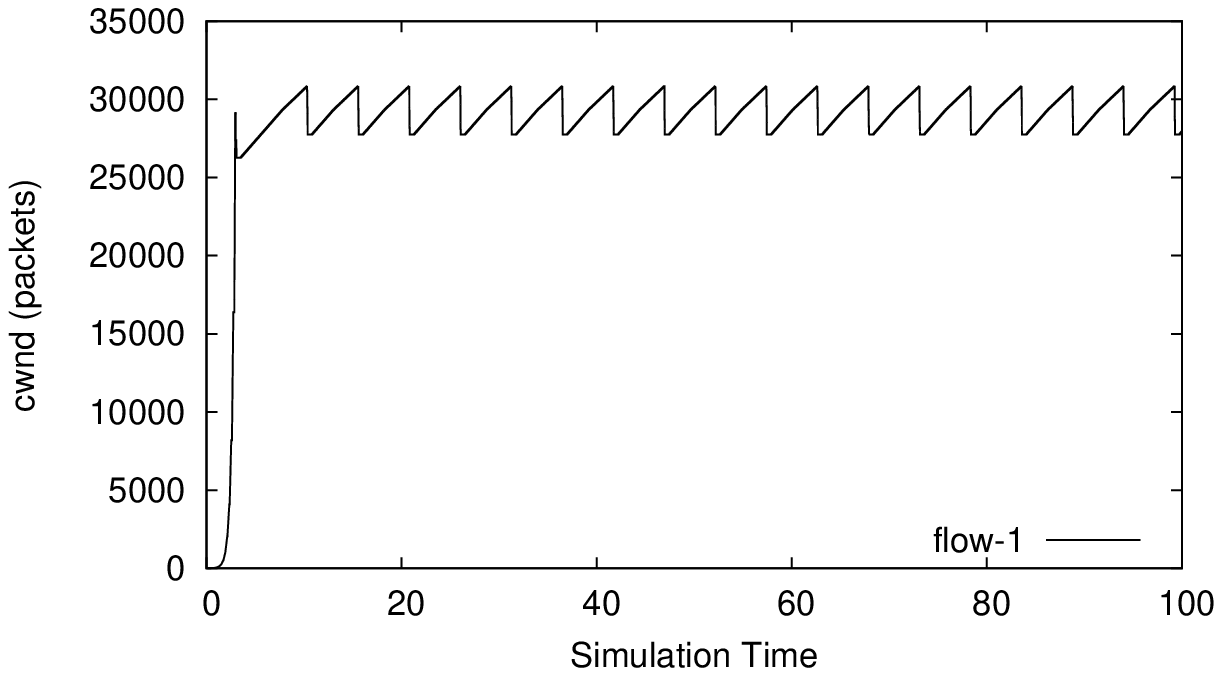}
			\label{fig:cwnd-agileld}
		}
		\subfigure[Cubic]
		{			
			\includegraphics[width=0.45\linewidth]{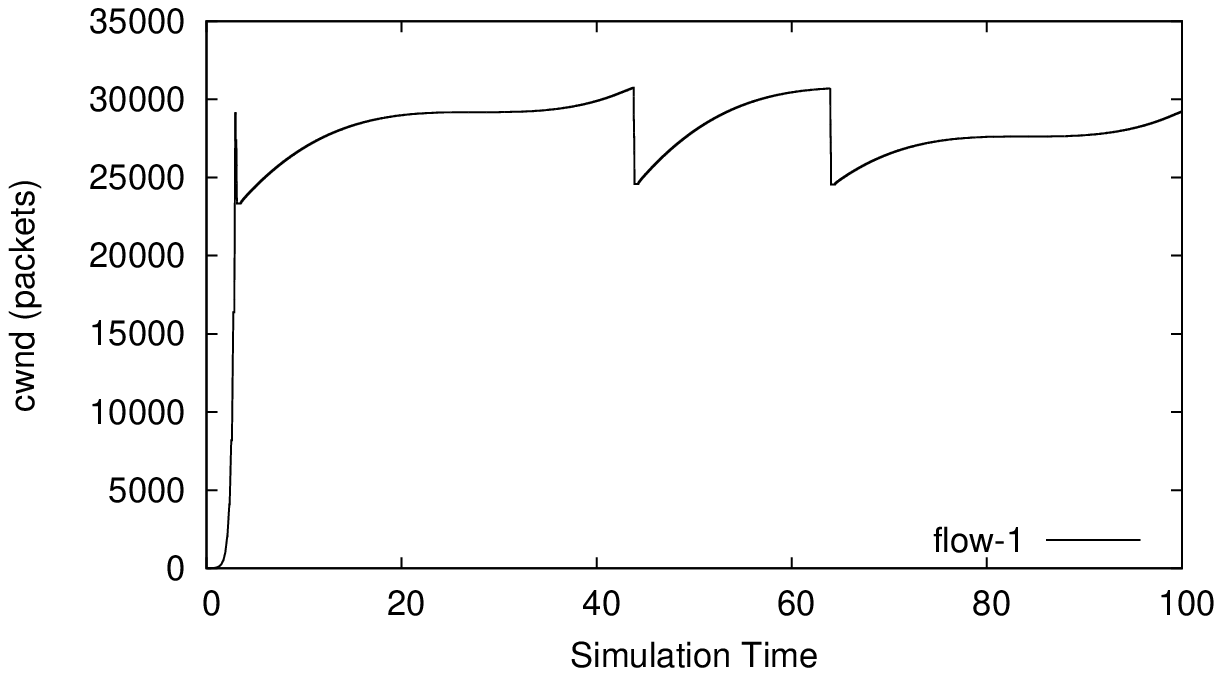}
			\label{fig:cwnd-cubic}
		}
		\subfigure[C-TCP]
		{			
			\includegraphics[width=0.45\linewidth]{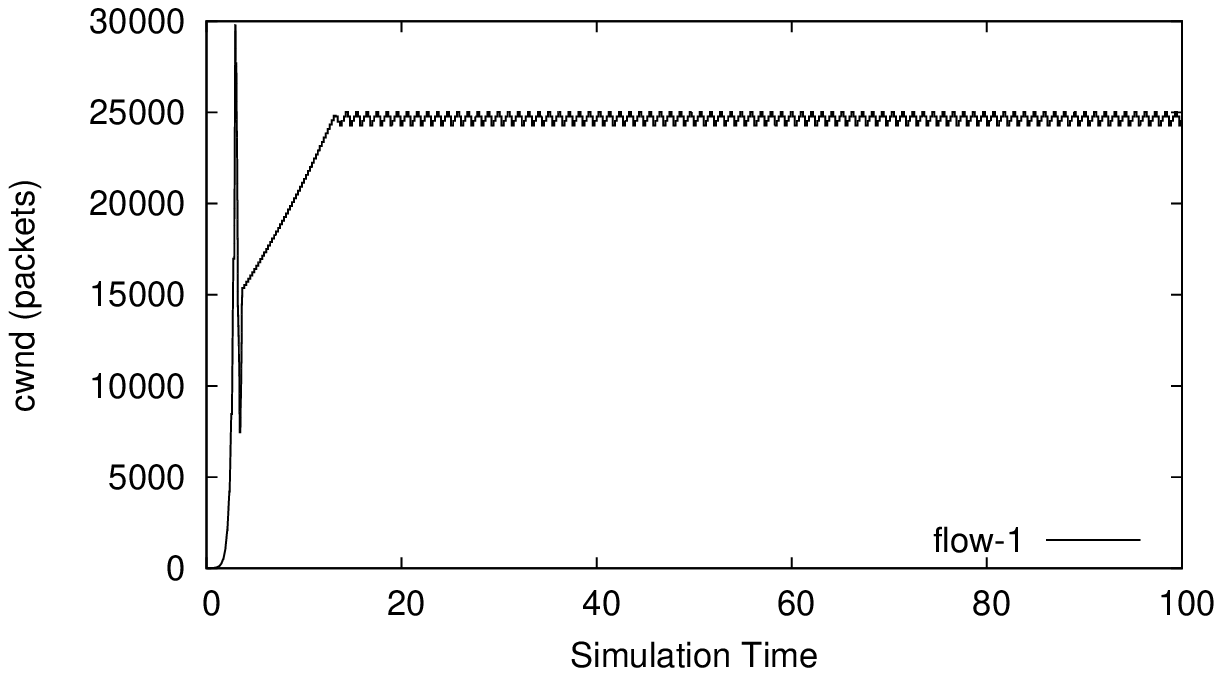}
			\label{fig:cwnd-compound}
		}
		\subfigure[Agile-SD]
		{	
			\includegraphics[width=0.45\linewidth]{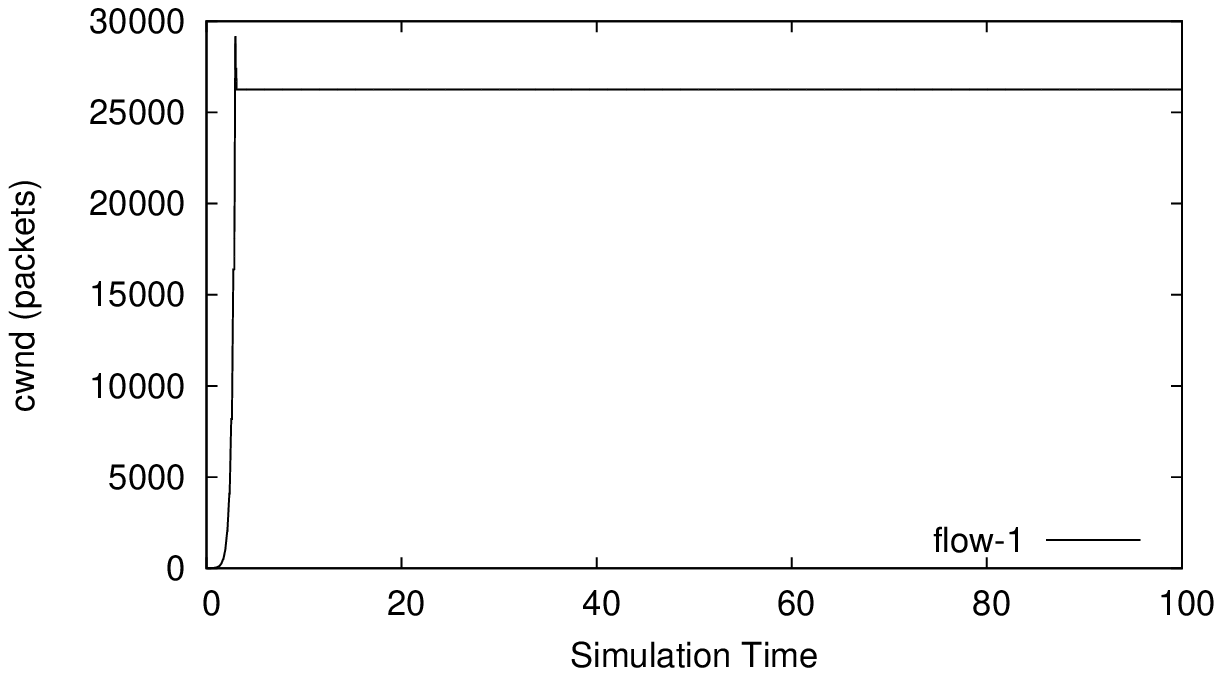}
			\label{fig:cwnd-agilesd}
		}
	\end{center}
	\caption{TCP congestion window evolution over single-flow scenario (buffer Size = 6400 packets, packet size = 1kbyte, loss rate = zero).}
	\label{fig:CWND}
\end{figure}

The evolution of \textit{cwnd} is the spirit of all CCAs, as it directly influences other performance metrics such as throughput, bandwidth utilization, loss ratio, and sharing fairness. Due to its unique behavior, Elastic-TCP expectedly presents faster \textit{cwnd} growth compared to Cubic, C-TCP, and Agile-SD, as shown in Figure \ref{fig:CWND}. This fast \textit{cwnd} growth allows Elastic-TCP to be an RTT-independent, which in turn shrinks its epoch time, where the faster the growth of \textit{cwnd} the shorter the epoch and vice versa. Indeed, shortening the epoch itself is not an aim, but it is only a way to increase the bandwidth utilization. By this approach, Elastic-TCP does not only increase the average throughput but also minimizes the loss ratio while maintaining the sharing fairness.

On one hand, Figure \ref{fig:CWND} shows the \textit{cwnd} evolution of the studied CCAs in the scenario of single-flow, where the faster increase is presented by the Elastic-TCP followed by Cubic, C-TCP, and Agile-SD. Clearly, the Elastic-TCP reaches roughly 31,000 packets in about 10 seconds, then it begins fluctuating to draw convex-up curves in very short epochs, as shown in Figure \ref{fig:cwnd-agileld}. With regard to Cubic, it reaches about 30,000 packets in 40 seconds, thereafter, it starts fluctuating to exhibit very long epochs due to its cubic function of the increase, as shown in Figure \ref{fig:cwnd-cubic}. While C-TCP does not exceed 25,000 packets, Agile-SD fixes its \textit{cwnd} to around 26,000 packets. Hence, it can be concluded that only the Elastic-TCP and Cubic have the ability to fully utilize the bandwidth in the ideal network, where the former is still better than the later by a difference of 1,000 packets (about 1Mbyte per RTT).

\begin{figure} [h!]
	\begin{center}
		\subfigure[Elastic-TCP]
		{
			\includegraphics[width=0.45\linewidth]{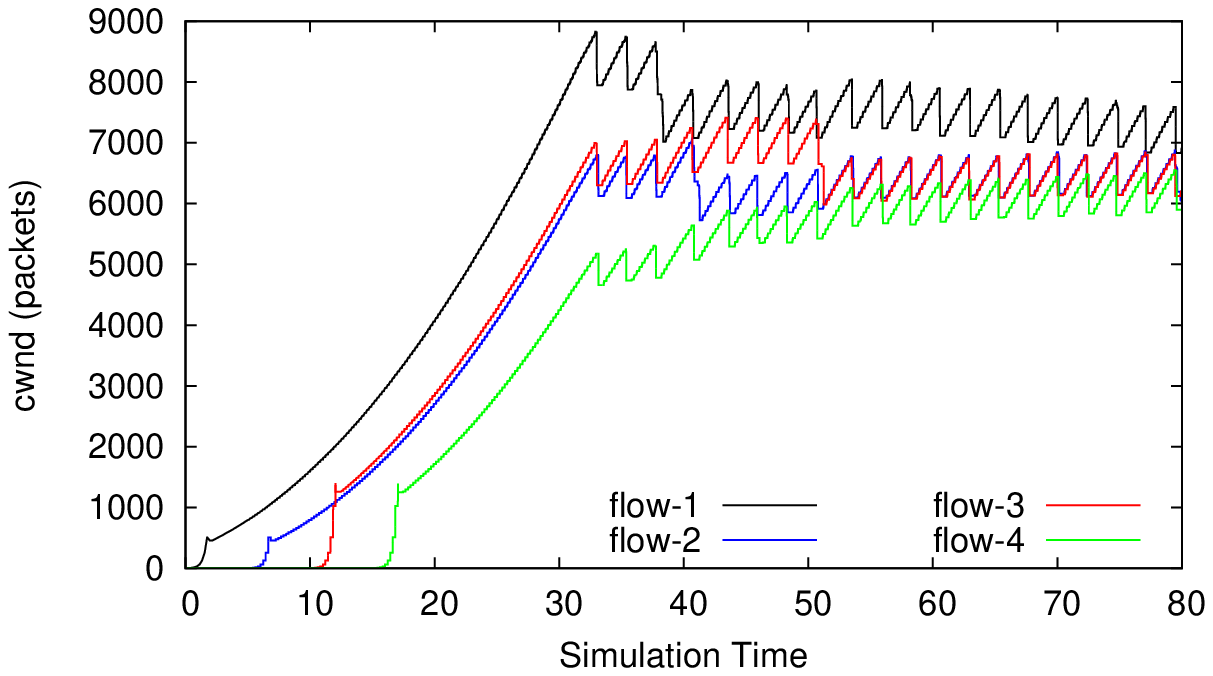}
			\label{fig:cwnd-agileld1}
		}
		\subfigure[Cubic]
		{			
			\includegraphics[width=0.45\linewidth]{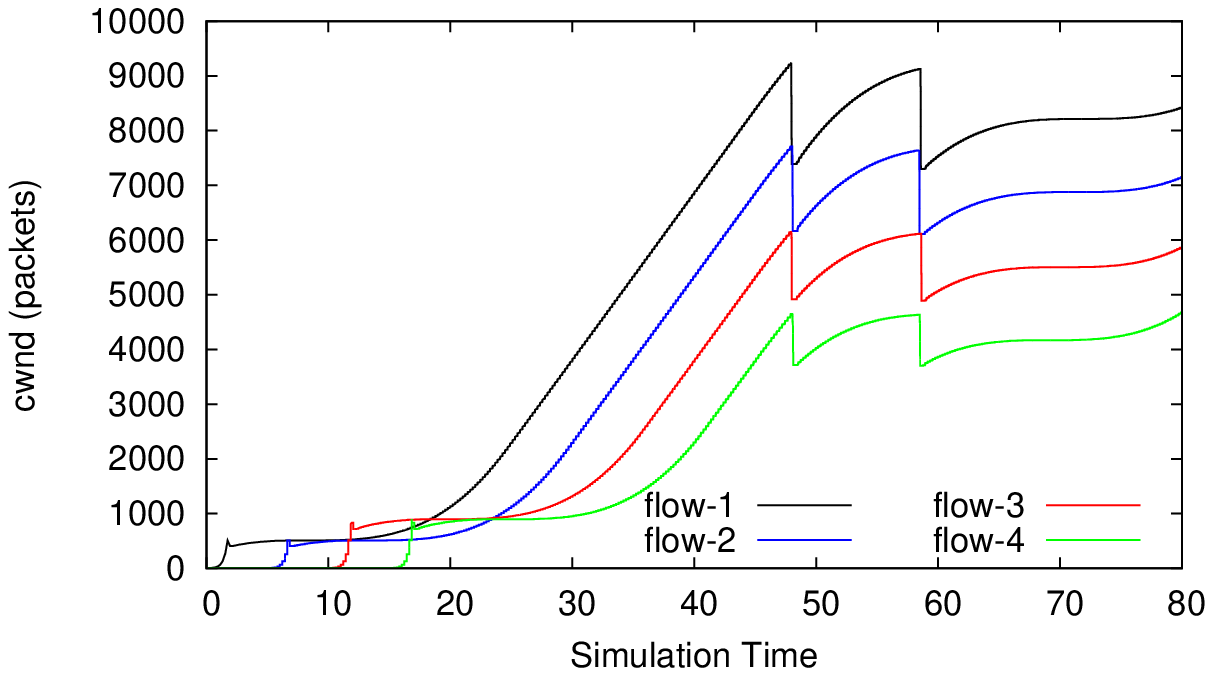}
			\label{fig:cwnd-cubic1}
		}
		\subfigure[C-TCP]
		{			
			\includegraphics[width=0.45\linewidth]{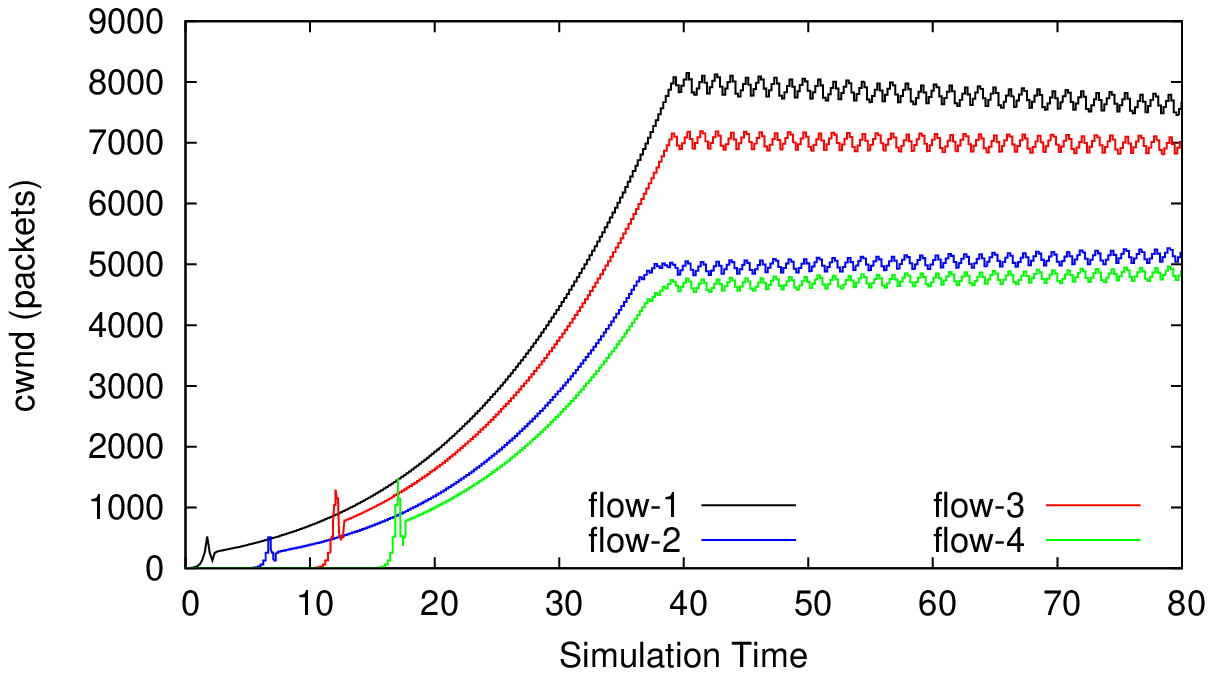}
			\label{fig:cwnd-compound1}
		}
		\subfigure[Agile-SD]
		{	
			\includegraphics[width=0.45\linewidth]{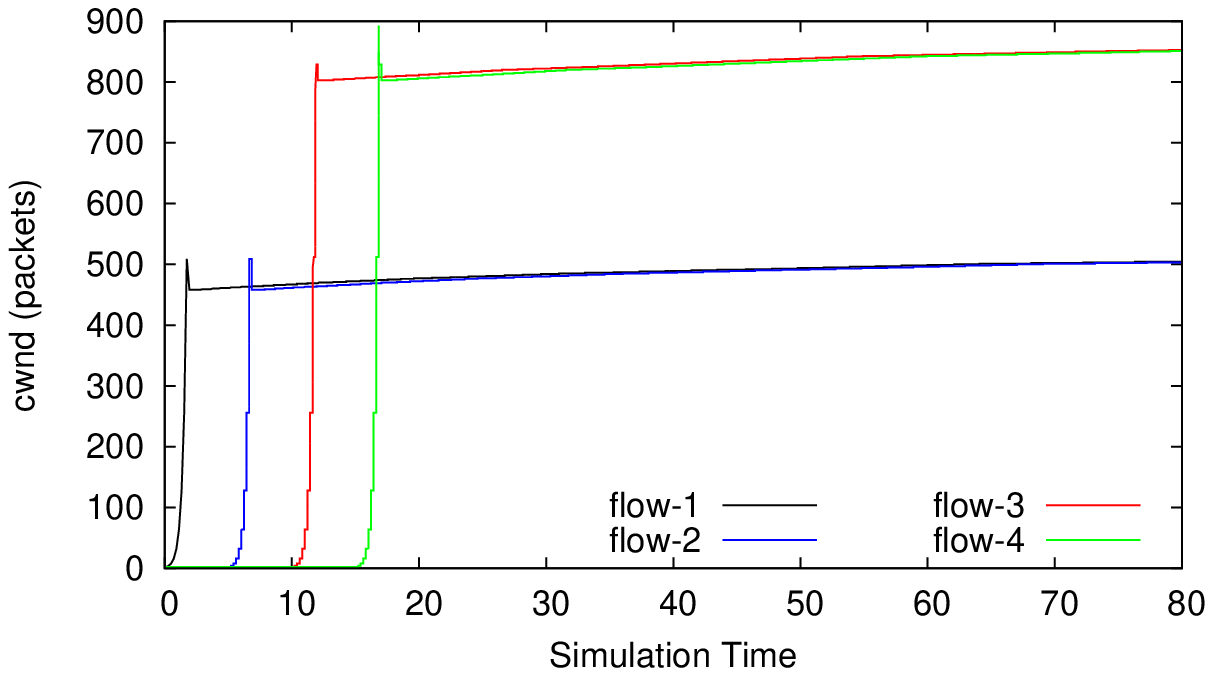}
			\label{fig:cwnd-agilesd1}
		}
	\end{center}
	\caption{TCP congestion window convergence in multi-flows scenario (buffer Size = 3200 packets, packet size = 1kbyte).}
	\label{fig:CWND2}
\end{figure}

On the other hand, Figure \ref{fig:CWND2} presents the \textit{cwnd} evolution of the studied CCAs in the scenario of multi-flows, with sequential flows establishments, to show the intra-fairness among these competing flows. Since the higher the convergence among concurrent flows the higher the intra-fairness, thus, the Elastic-TCP shows the highest intra-fairness level followed by C-TCP, Cubic and Agile-SD, and also Elastic-TCP shows higher utilization. 

More specifically, Elastic-TCP flows start converging with each other in around 35 seconds and they finish with a very high level of intra-fairness, while C-TCP flows start their convergence in about 40 seconds, but they finish with slightly lower intra-fairness than the former. As for Cubic, the flows start converging very slowly in 50 seconds and they give a moderate level of intra-fairness. Regarding Agile-SD, it exhibits a low level of fairness and very low bandwidth utilization with \textit{cwnd} not more than 900 packets, while the \textit{cwnd} of the other CCAs varies from 4,000 to 9,000 packets.

\begin{figure} [h!]
	\begin{center}
		\subfigure[Loss rate = 0]
		{
			\includegraphics[width=0.45\linewidth]{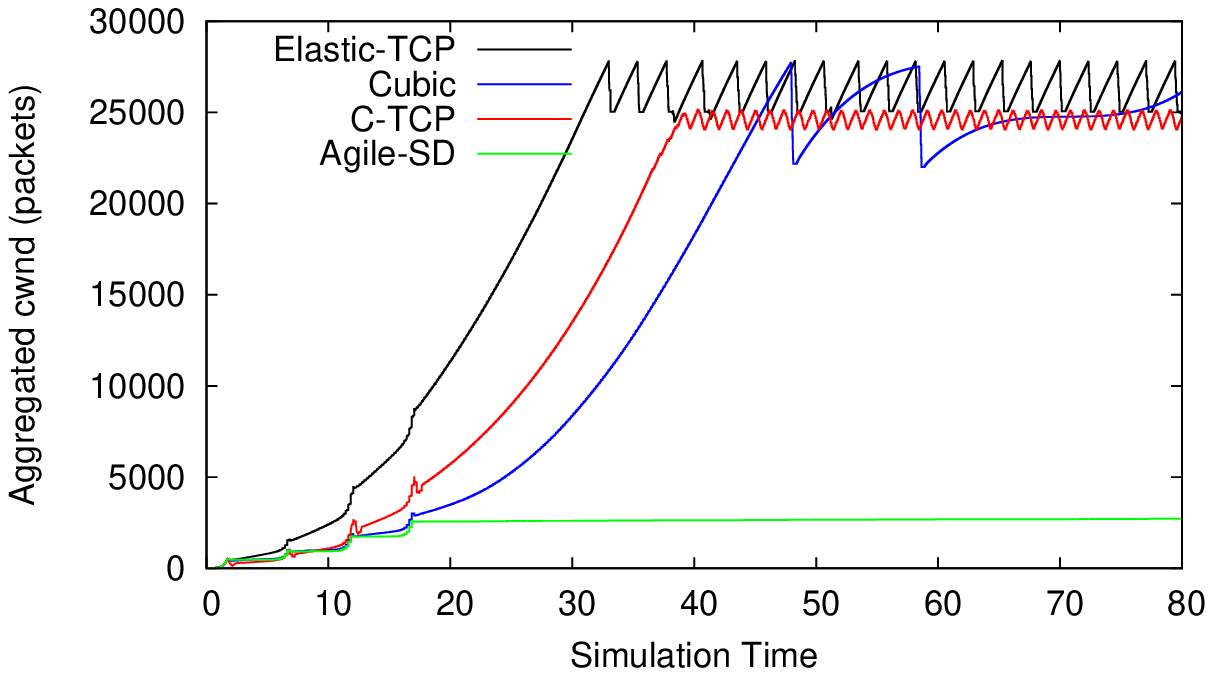}
			\label{fig:aggcwnd-zero}
		}
		\subfigure[loss rate = $10^{-5}$]
		{			
			\includegraphics[width=0.45\linewidth]{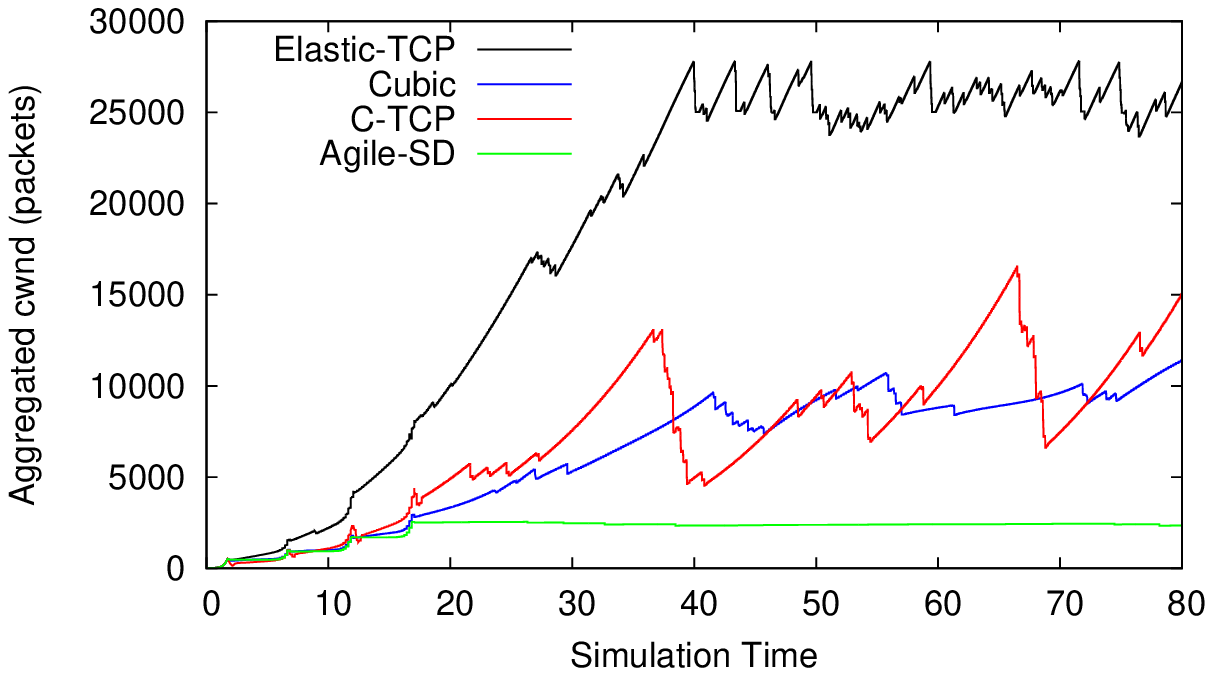}
			\label{fig:aggcwnd-10-5}
		}
	\end{center}
	\caption{TCP aggregated \textit{cwnd} in multi-flows scenario (buffer Size = 3200 packets, packet size = 1kbyte).}
	\label{fig:CWND3}
\end{figure}

Figure \ref{fig:CWND3} shows a comparison between the studied CCAs in terms of \textit{cwnd} evolution. It shows the average \textit{cwnd} of four concurrent flows for each CCA in the case of zero PER and $10^{-5}$ PER. From Figure \ref{fig:aggcwnd-zero}, it is clear that Elastic-TCP reaches the maximum \textit{cwnd} earlier than C-TCP and Cubic, while Agile-SD is not able to reach reasonable \textit{cwnd} value since it is not designed for high-BDP networks. Moreover, C-TCP and Cubic show lower \textit{cwnd} than Elastic-TCP even after they reach their steady states. In Figure \ref{fig:aggcwnd-10-5}, Cubic and C-TCP show high sensitivity to packet loss and both degrade their \textit{cwnd} to less than 50\%, while Elastic-TCP shows very low sensitivity to packet loss which allows it to maintain a high level of performance.

\subsubsection{The average throughput}
\begin{figure} [h!]
	\begin{center}
		\subfigure[$zero$ PER.] 
		{
			\includegraphics[width=0.45\linewidth]{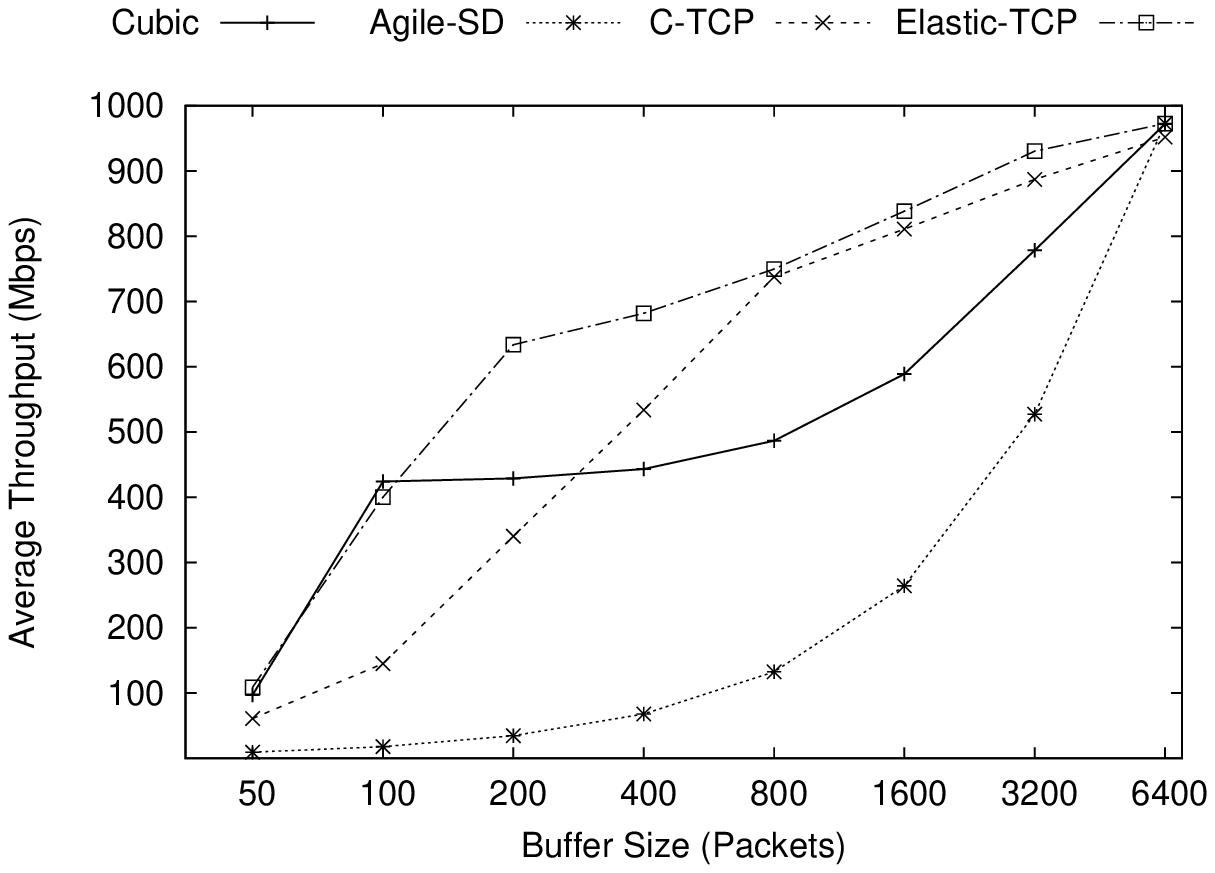}
			\label{fig:single-0per-throughput}
		}
		\subfigure[$10^{-5}$ PER.]
		{
			\includegraphics[width=0.45\linewidth]{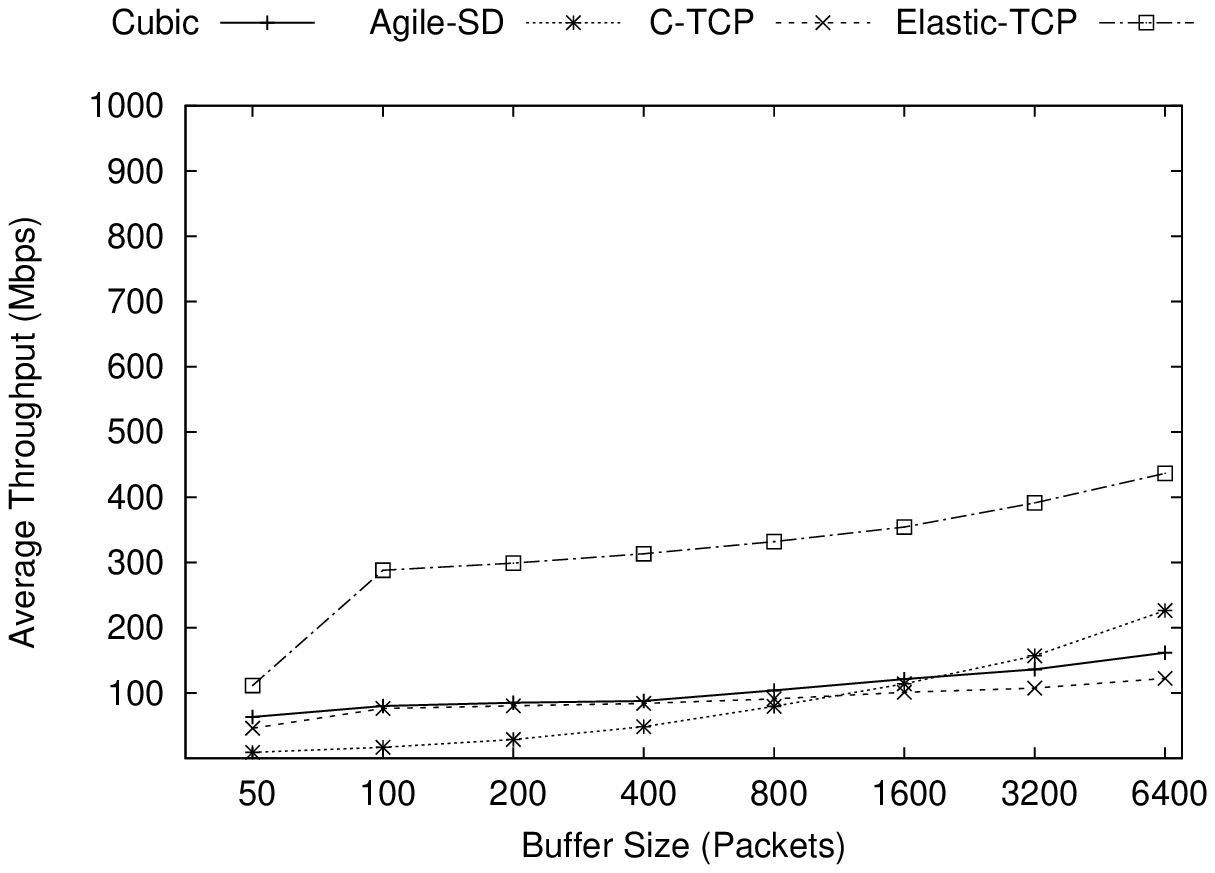}
			\label{fig:single-2per-throughput}
		}
		\subfigure[$10^{-4}$ PER.]
		{
			\includegraphics[width=0.45\linewidth]{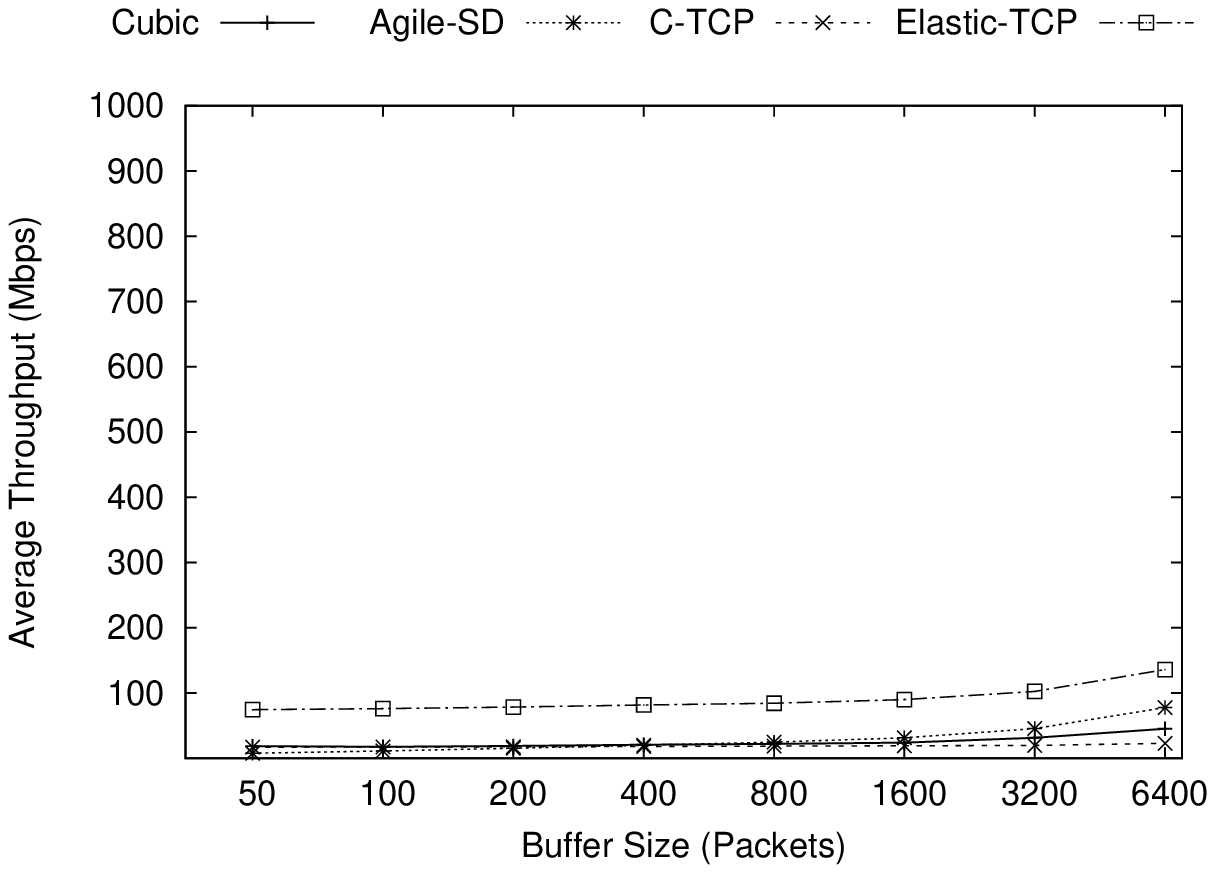}
			\label{fig:single-3per-throughput}
		}
	\end{center}
	\caption{The single flow scenario: the average throughput against buffer size.}
	\label{fig:sync-throughput1}
\end{figure}

The single-flow scenario shows an ideal congestion-free network to study the capability of TCP CCAs on fully utilizing the available bandwidth. The proposed CCA shows slight enhancement on average throughput compared to other CCAs due to the fast increase of its \textit{cwnd} resulted by its unique mechanism of WWF, as shown in Figure \ref{fig:single-0per-throughput}. Moreover, the Elastic-TCP shows more sustainability in presence of PER compared to other CCAs, as shown in figures \ref{fig:single-2per-throughput} and \ref{fig:single-3per-throughput}, where Cubic, C-TCP, and Agile-SD are highly influenced by the PER. In general, the Elastic-TCP outperforms other CCAs in terms of throughput in most cases even in harsh network environments where PER is high. This clearly enhances the bandwidth utilization by up to 22\% in some scenarios.

\begin{figure} [h!]
	\begin{center}
		\subfigure[$zero$ PER.] 
		{
			\includegraphics[width=0.45\linewidth]{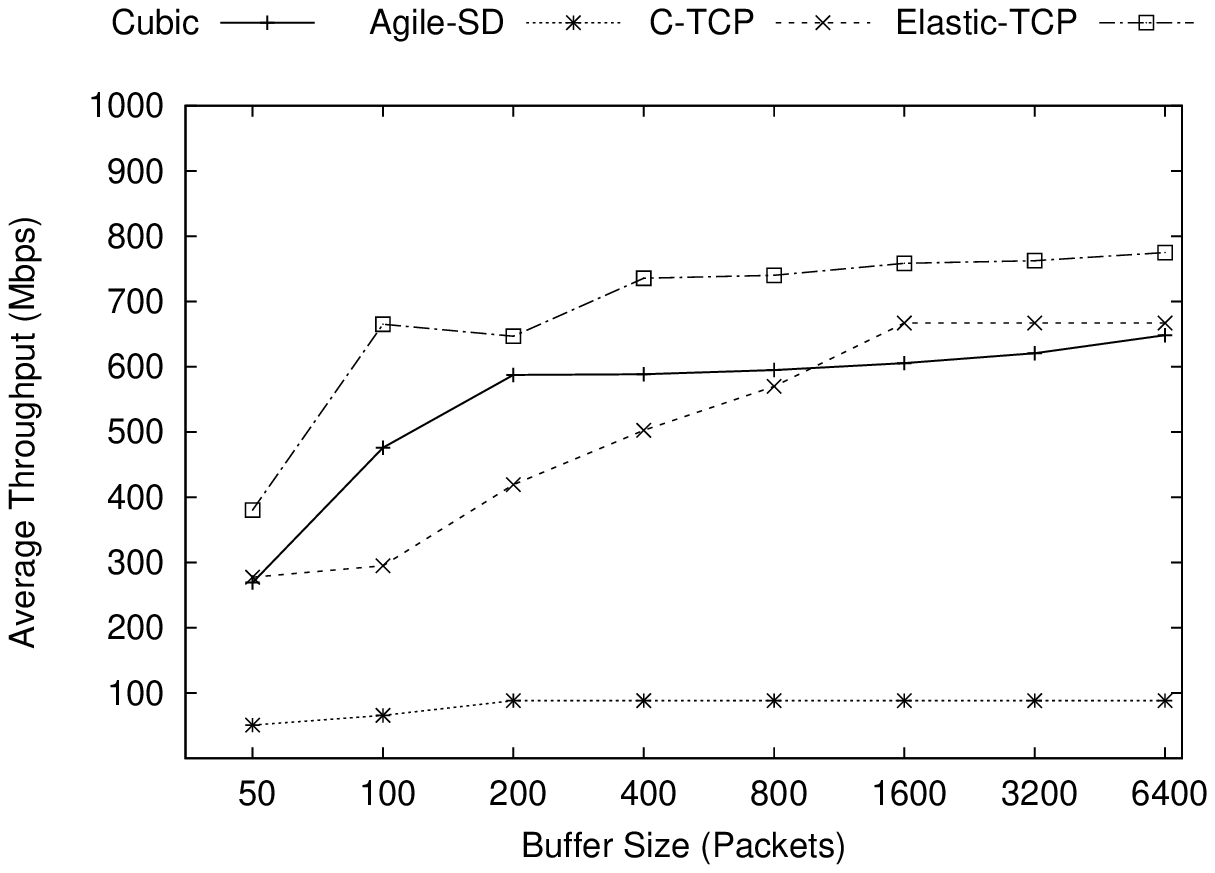}
			\label{fig:seq-0per-throughput}
		}
		\subfigure[$10^{-5}$ PER.]
		{
			\includegraphics[width=0.45\linewidth]{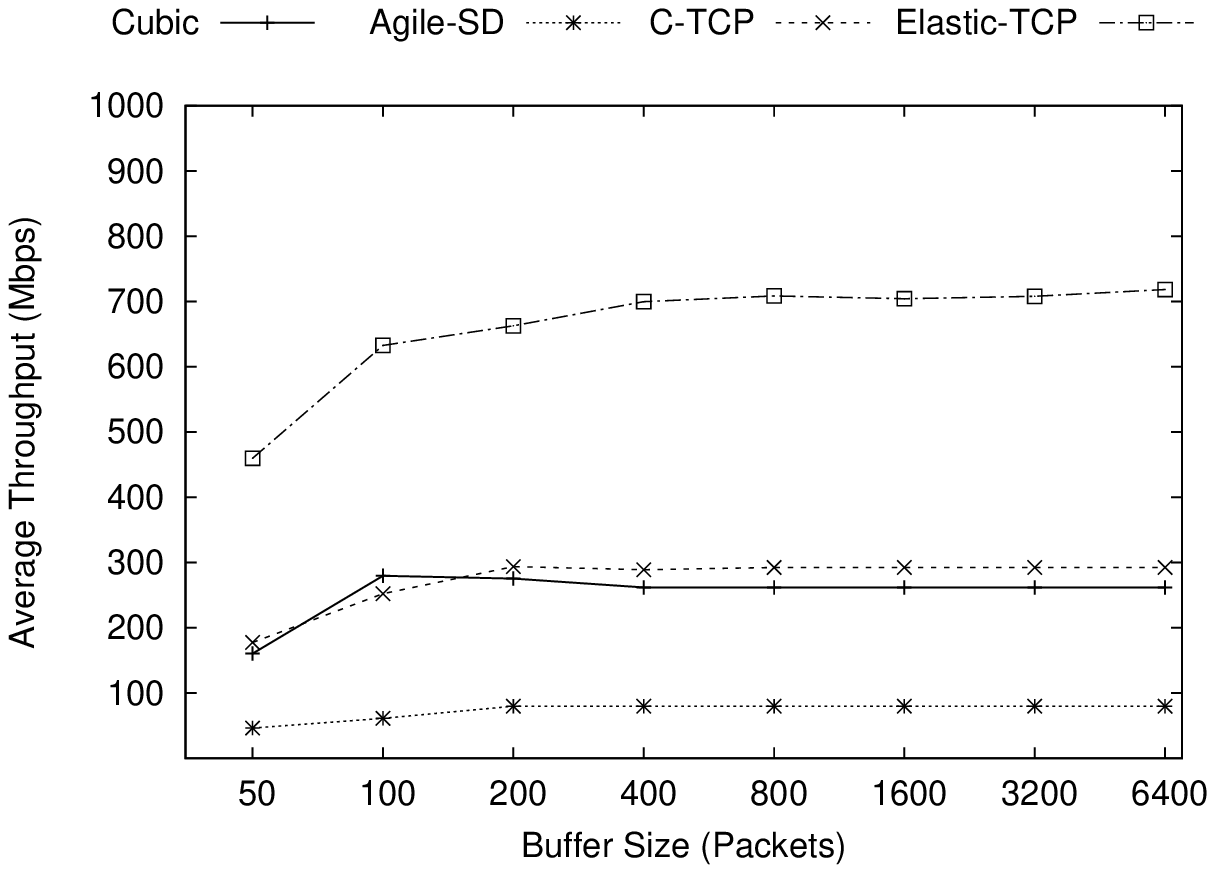}
			\label{fig:seq-2per-throughput}
		}
		\subfigure[$10^{-4}$  PER.]
		{
			\includegraphics[width=0.45\linewidth]{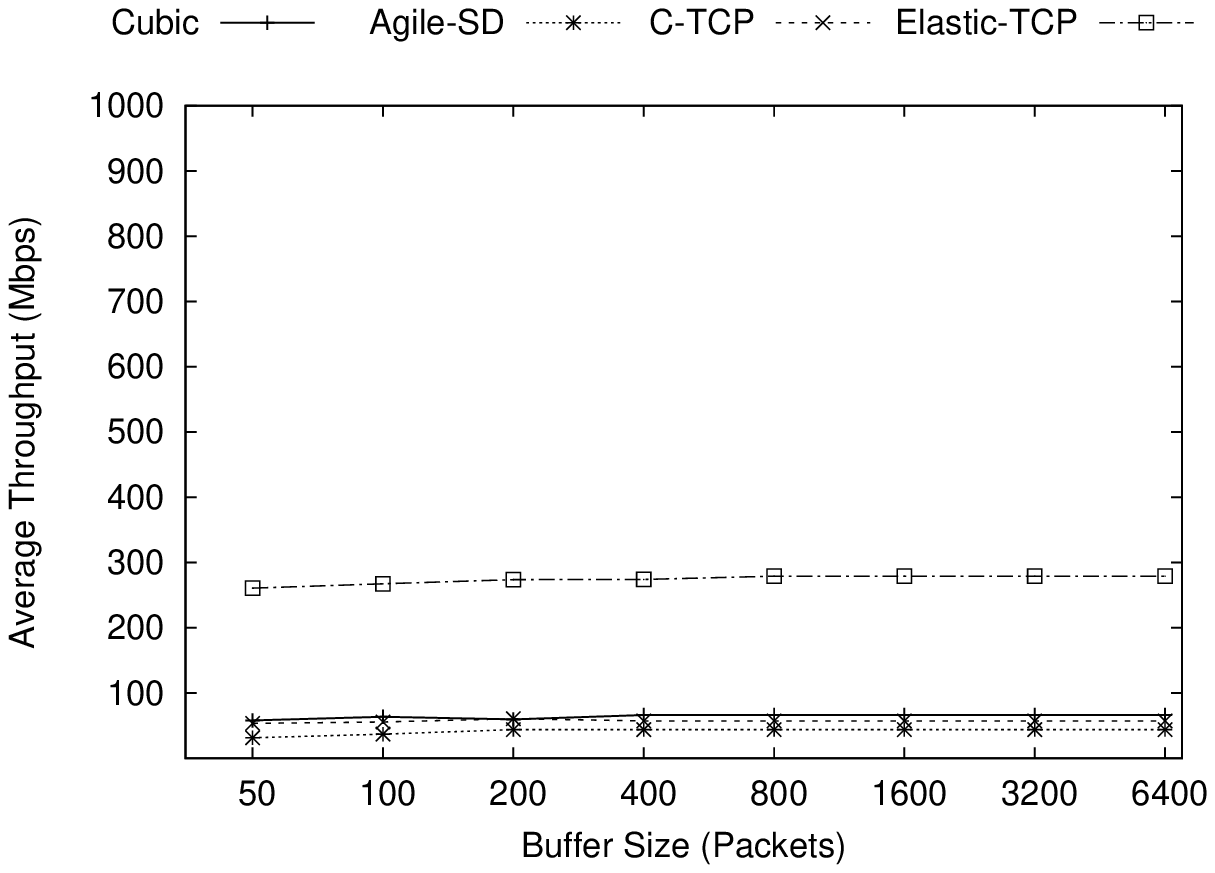}
			\label{fig:seq-3per-throughput}
		}
	\end{center}
	\caption{Sequential multiple-flows scenario: average throughput vs. buffer size.}
	\label{fig:sync-throughput2}
\end{figure}

In the second scenario, figures \ref{fig:seq-0per-throughput}, \ref{fig:seq-2per-throughput} and \ref{fig:seq-3per-throughput} show that the Elastic-TCP achieves better throughput compared to other CCAs, even with small buffer size and high PER, which enhances the bandwidth utilization up to 40\%.

\begin{figure} [h!]
	\begin{center}
		\subfigure[$zero$ PER.] 
		{
			\includegraphics[width=0.45\linewidth]{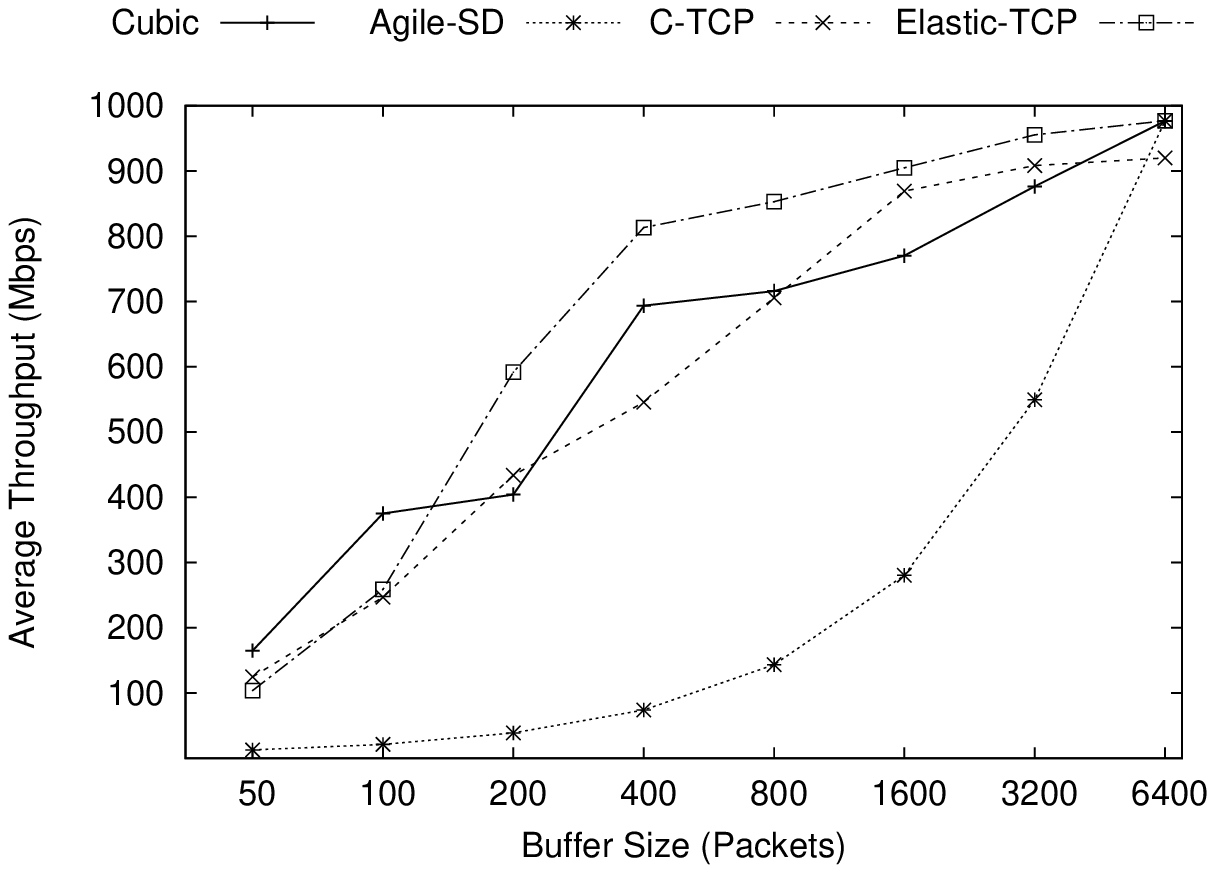}
			\label{fig:sync-0per-throughput}
		}
		\subfigure[$10^{-5}$ PER.] 
		{
			\includegraphics[width=0.45\linewidth]{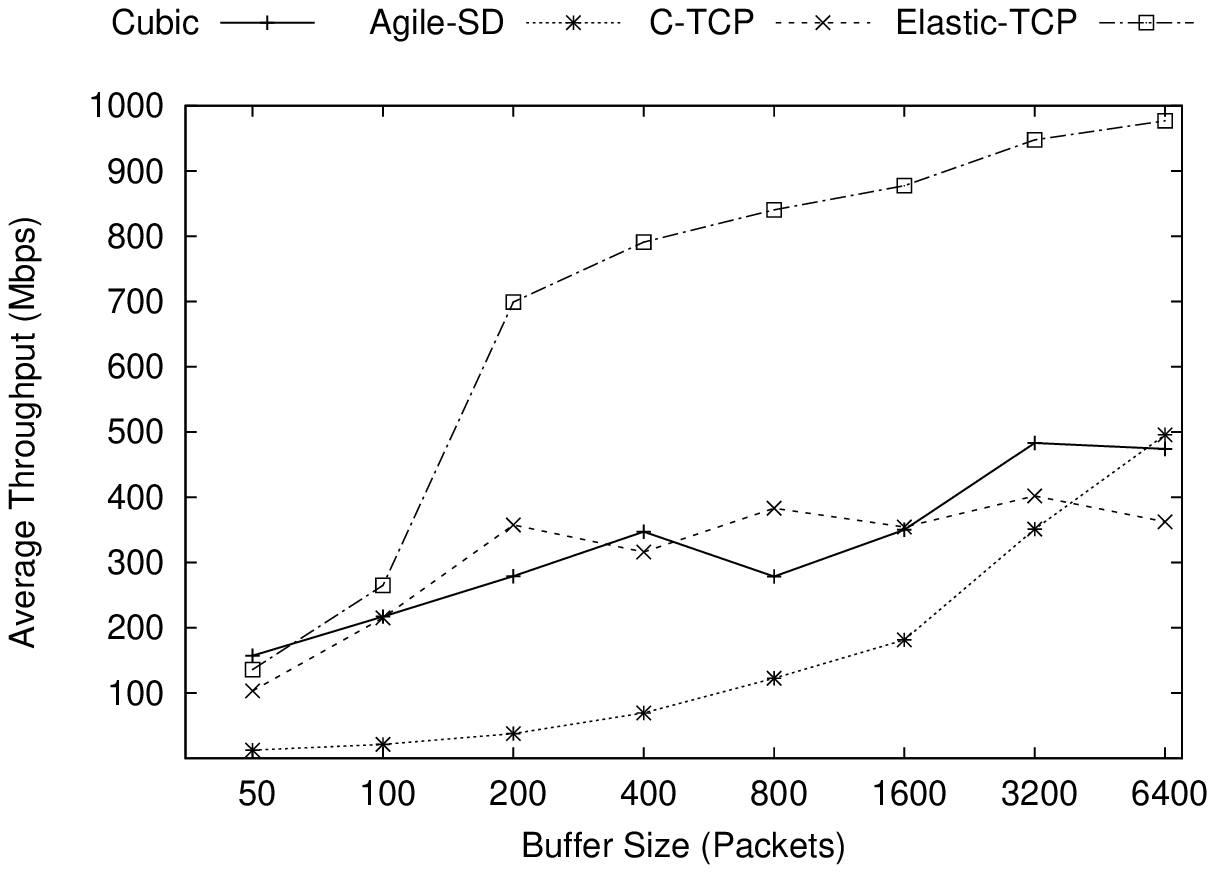}
			\label{fig:sync-2per-throughput}
		}
		\subfigure[$10^{-4}$ PER.] 
		{
			\includegraphics[width=0.45\linewidth]{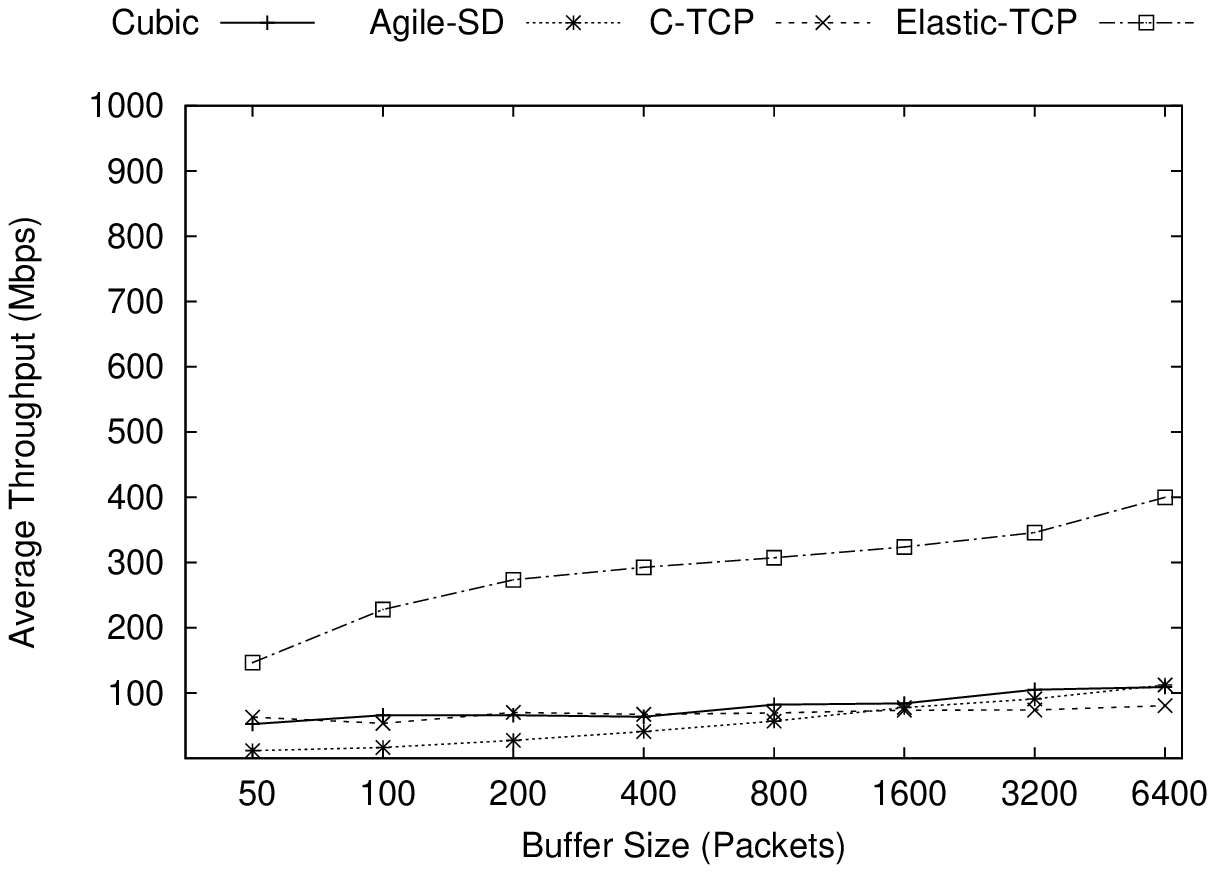}
			\label{fig:sync-3per-throughput}
		}
	\end{center}
	\caption{Synchronous multiple-flows scenario: average throughput vs. buffer size.}
	\label{fig:sync-throughput3}
\end{figure}

In the synchronous multiple-flows scenario, the Elastic-TCP also outperforms the other CCAs in most cases, especially with high PER and it significantly achieves up to 50\% of improvement in some cases, as shown in figures \ref{fig:sync-0per-throughput}, \ref{fig:sync-2per-throughput} and \ref{fig:sync-3per-throughput}.

\subsubsection{The loss ratio}
Fundamentally, TCP aims at maximizing the throughput while minimizing the loss ratio. Thus, in all scenarios, the new Elastic-TCP along with the studied CCAs produce very trivial loss ratios, which is not more than 0.5\%, as shown in Table \ref{fig:lr}, where the rest of results have no much difference.

\begin{table}[h!]
	\caption{Loss ratio vs. buffer size: synchronous multi-flows scenario, $zero$ PER.}
	\begin{center}
		{\footnotesize \begin{tabular}{c|cccc}
				\hline
				\multirow{2}{*}{Buffer}			& \multicolumn{4}{c}{Loss ratio} \\ \cline{2-5} 
												& Cubic    &   C-TCP  &  Agile-SD &  Elastic-TCP \\ \hline
				\rowcolor[HTML]{EFEFEF} 50 	   	& 0.006840 & 0.036343 &  0.058301 &   0.009872 \\
										100    	& 0.004418 & 0.031290 &  0.060696 &   0.003612 \\
				\rowcolor[HTML]{EFEFEF} 200   	& 0.006269 & 0.017994 &  0.062253 &   0.024834 \\
										400    	& 0.010915 & 0.024560 &  0.063563 &   0.028342 \\
				\rowcolor[HTML]{EFEFEF} 800    	& 0.018782 & 0.012103 &  0.065065 &   0.035166 \\
										1600   	& 0.030127 & 0.022083 &  0.065139 &   0.045517 \\
				\rowcolor[HTML]{EFEFEF} 3200   	& 0.044965 & 0.040465 &  0.065239 &   0.063763 \\
										6400   	& 0.071371 & 0.075520 &  0.070707 &   0.094607 \\ \hline
		\end{tabular}}
		\label{fig:lr}
	\end{center} \vspace{-0.5cm}
\end{table}

\subsubsection{The fairness}
\begin{figure} [h!]
	\begin{center}
		\subfigure[Intra-fairness index against buffer size: synchronous multi-flows scenario, $10^{-5}$ PER.] 
		{
			\includegraphics[width=0.45\linewidth]{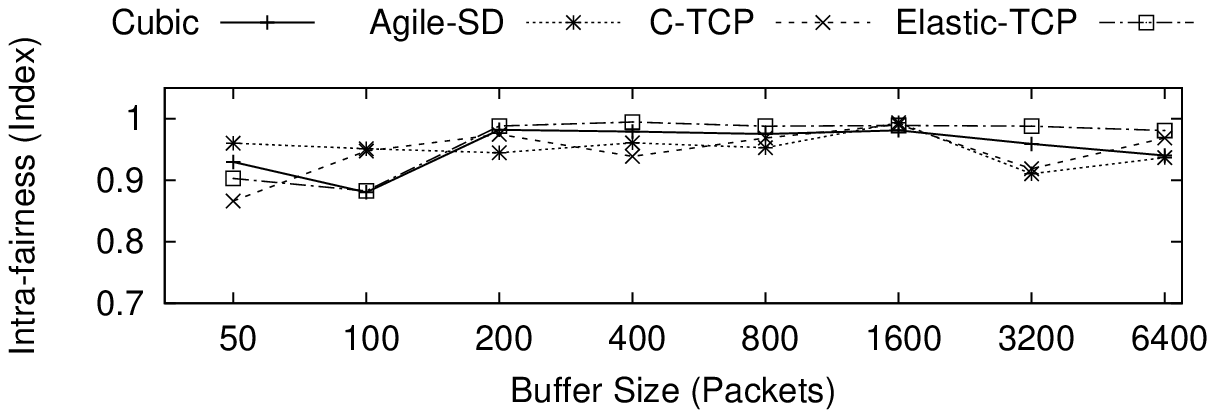}
			\label{fig:intra}
		}
		\subfigure[RTT-fairness index against buffer size: synchronous multi-flows scenario, $10^{-5}$ PER.] 
		{
			\includegraphics[width=0.45\linewidth]{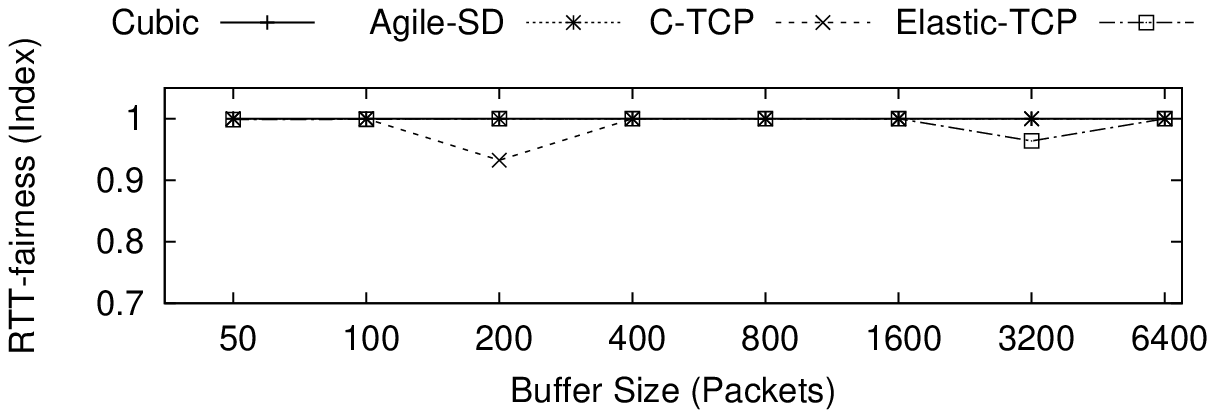}
			\label{fig:rtt}
		}
		\subfigure[Inter-fairness among the studied CCAs.] 
		{
			\includegraphics[angle=-90,width=0.45\linewidth]{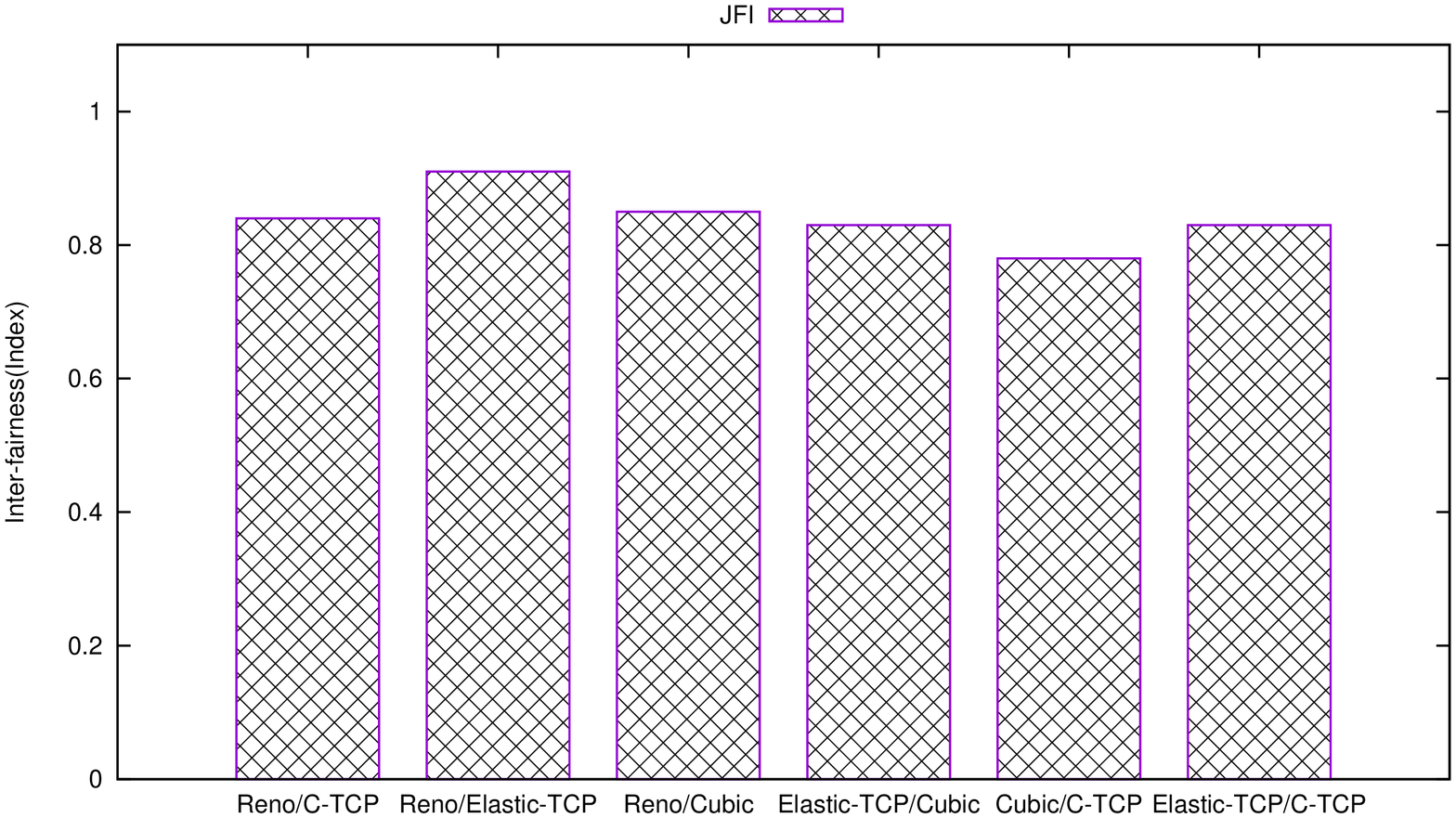}
			\label{fig:inter}
		}
		\caption{The fairness measurements.}
	\end{center}
\end{figure}

Simulation results show that all examined CCAs attain similar intra-fairness and RTT-fairness. However, thanks to the weighting function that enabled the Elastic-TCP to achieve slightly higher fairness index than other CCAs, especially in high PER and small buffer cases. Due to the trivial difference in fairness results among examined CCAs, figures \ref{fig:intra} and \ref{fig:rtt} were chosen to show samples of intra-fairness and RTT-fairness, respectively.

Moreover, the inter-fairness of the examined CCAs against standard NewReno is measured in an individual experiment using the topology shown in Figure \ref{fig:topology}, where the result of this metric is shown in Figure \ref{fig:inter}. For inter-fairness to NewReno, the Elastic-TCP achieves the highest score, which is around 91\%, while Cubic-TCP and C-TCP achieve about 85\% inter-fairness measurement. With regard to inter-fairness to Cubic-TCP, the Elastic-TCP and NewReno achieve the highest index which is about 84\% while C-TCP achieves only 78\%. For inter-fairness to C-TCP, both Elastic-TCP and NewReno attain about 85\% while Cubic-TCP attains only 78\%. In fact, the Elastic-TCP achieves a high level of inter-fairness to other standard CCAs due to its unique functionality of WWF.

\section{Performance Evaluation of Elastic-TCP using Testbed}
\label{PET}

The proposed Elastic-TCP is compiled into the Linux kernel, version 4.9 using openSUSE Leap 42.2, to carry out the testbed experiment, in order to show the performance of Elastic-TCP in the real environment. Since Elastic-TCP is designed for long-distance networks, we used the Linux-based NetEm to emulate the delay and to control the buffer size.

\subsection{Testbed Setup}
\begin{figure}[h!]
	\centering
	\includegraphics[trim={2mm 2 2 2},clip,width=0.7\linewidth]{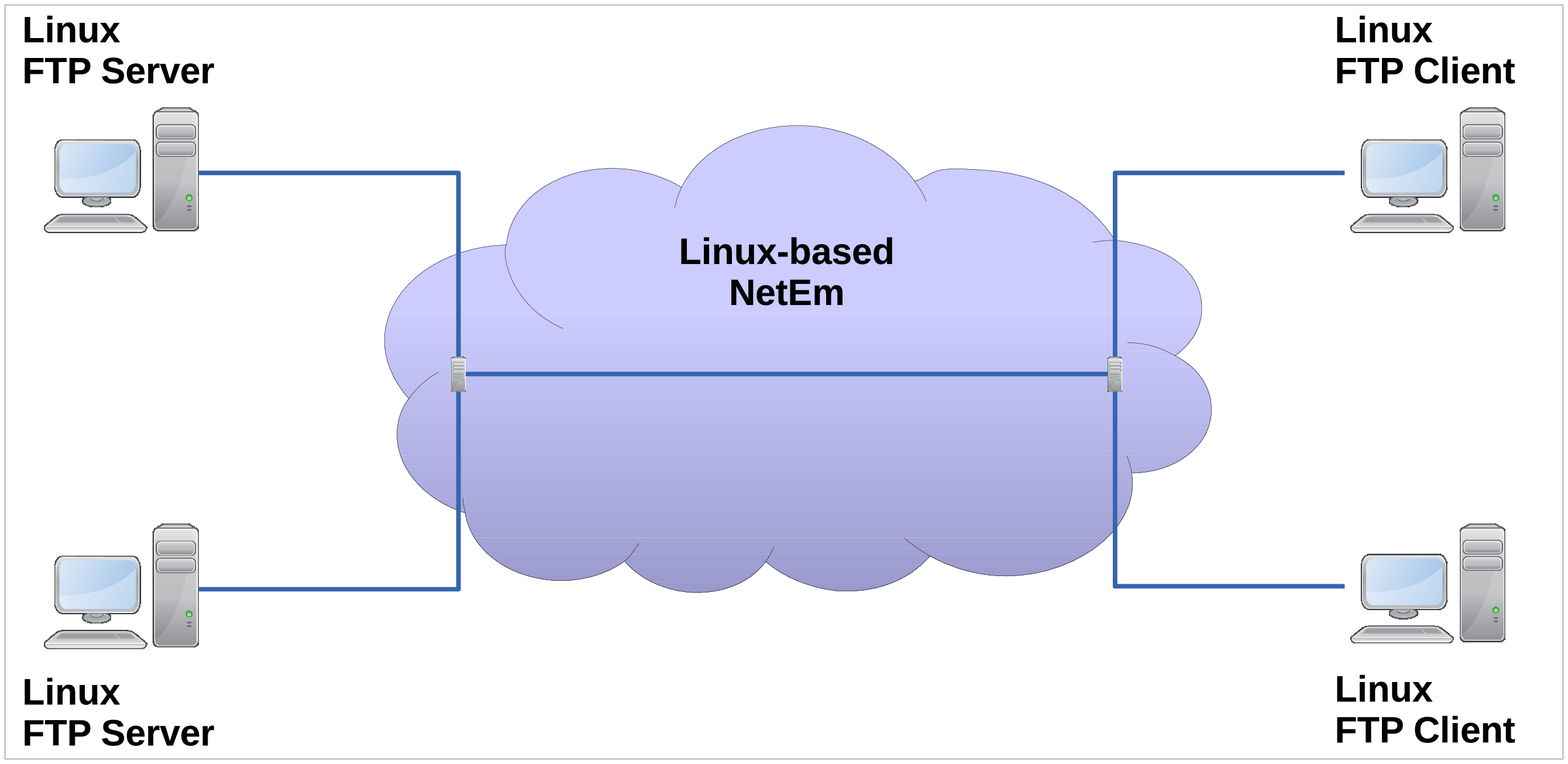}
	\caption{NetEm-based Testbed Topology.}
	\label{fig:testbed}
\end{figure}

A testbed of single dumbbell topology is built in our \mbox{laboratory} using real PCs connected to each other through 1Gbps wired links, as shown in Figure \ref{fig:testbed}. In order to build this network topology, we installed Linux openSUSE 42.2 Leap over all servers and clients. Thereafter, we implemented our Elastic-TCP module into the Linux kernel over all servers and clients. In order to evaluate the tested CCAs, we transfer large files from the clients to the servers simultaneously, while the network traffic is monitored using TCPdump. As for NetEm, it is configured at all end-systems to provide 100ms round-trip time for all links. The experiment is repeated 30 times for every buffer size scenario, where the buffer size is varied from 50 to 12,500 packets. For the average throughput, the Standard Deviation (SD) with 95\% Confident Interval (CI) and Standard Error have been calculated for every 30 runs for every set of parameter setup. 

Moreover, the studied CCAs are evaluated over two scenarios, single-flow and multiple-flows scenario. In order to make our performance evaluation up to date, we included the TCP-BBR to our comparison. TCP-BBR is recently developed by Google and currently becomes the most promising candidate to replace current congestion control protocols in the upcoming 4.9 Linux kernel. Table \ref{params2} shows the testbed parameters' setup and configuration.

\begin{table}[h!]
	\caption{Testbed parameters setup and configuration.}
	\begin{center}
		\begin{tabular}{p{2.6cm}p{4.5cm}}
			\hline					Parameter				&	Value (s)							\\ \hline
			\rowcolor[HTML]{EFEFEF} TCP CCAs				&	Cubic, C-TCP, TCP-BBR, Elastic-TCP	\\ 
									Link capacity			&	1Gbps for all links					\\ 
			\rowcolor[HTML]{EFEFEF} Two-way delay			&	100ms for all links					\\ 
									Buffer size				&	from 50 to 12500 packets			\\ 
			\rowcolor[HTML]{EFEFEF} Packet size				&	1500 bytes							\\ 
									Queuing Algo	 		&	Drop Tail							\\ 
			\rowcolor[HTML]{EFEFEF} Traffic type			&	FTP									\\ 
									Transfered file size	&	5.1GB								\\ 
			\rowcolor[HTML]{EFEFEF}	Runs for Each Scenario 	&	30 times							\\ \hline
		\end{tabular}
		\label{params2}
	\end{center} \vspace{-0.5cm}
\end{table}

\subsection{Testbed Results and Discussion}
This subsection analytically discusses the testbed results and shows the average throughput, loss ratio, and fairness measurements in order to show the impact of long-delay and buffer size on the overall performance.
\subsubsection{The average throughput}
As shown in Figure \ref{fig:14}, Elastic-TCP achieves higher average throughput compared to other TCP CCAs as a result of its fast \textit{cwnd} growth resulted by its unique WWF mechanism. The Elastic-TCP performs better than the compared CCAs in most cases, particularly when the applied buffer size is small. In single-flow scenario, the Elastic-TCP improves the average throughput by up to 14\% over TCP-BBR, up to 13\% over Cubic and up to 154\% over C-TCP. In multiple-flows scenario, Figure \ref{fig:17} shows that the Elastic-TCP outperforms the compared CCAs, in terms of average throughput, in many cases, especially when the applied buffer size is small. Briefly, it enhances the average throughput by up to 23\% over TCP-BBR, up to 14\% over Cubic and up to 81\% over C-TCP.

\subsubsection{The loss ratio}
\begin{figure} [h!]
	\centering
	\begin{center}
		\subfigure[Average throughput \& SD with CI 95\%.] 
		{
			\includegraphics[width=0.45\linewidth]{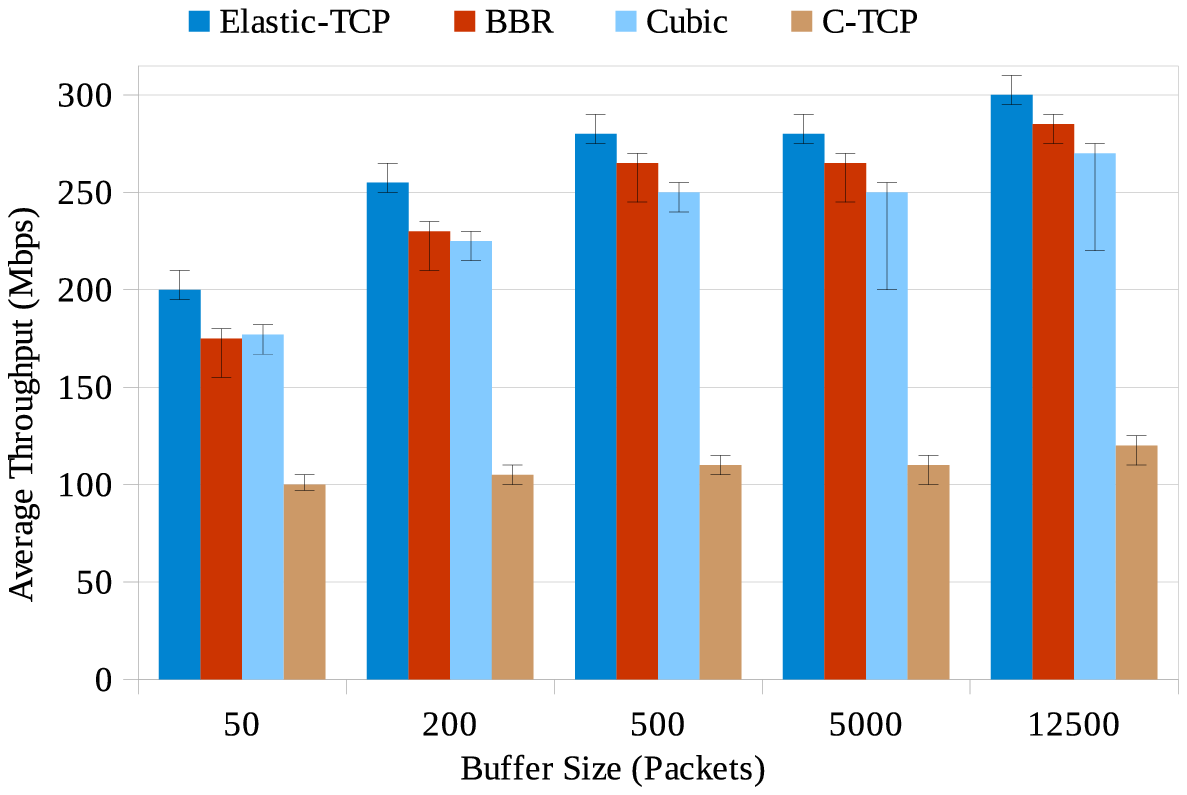}
			\label{fig:14}
		}
		\subfigure[Loss ratio.] 
		{
			\includegraphics[width=0.45\linewidth]{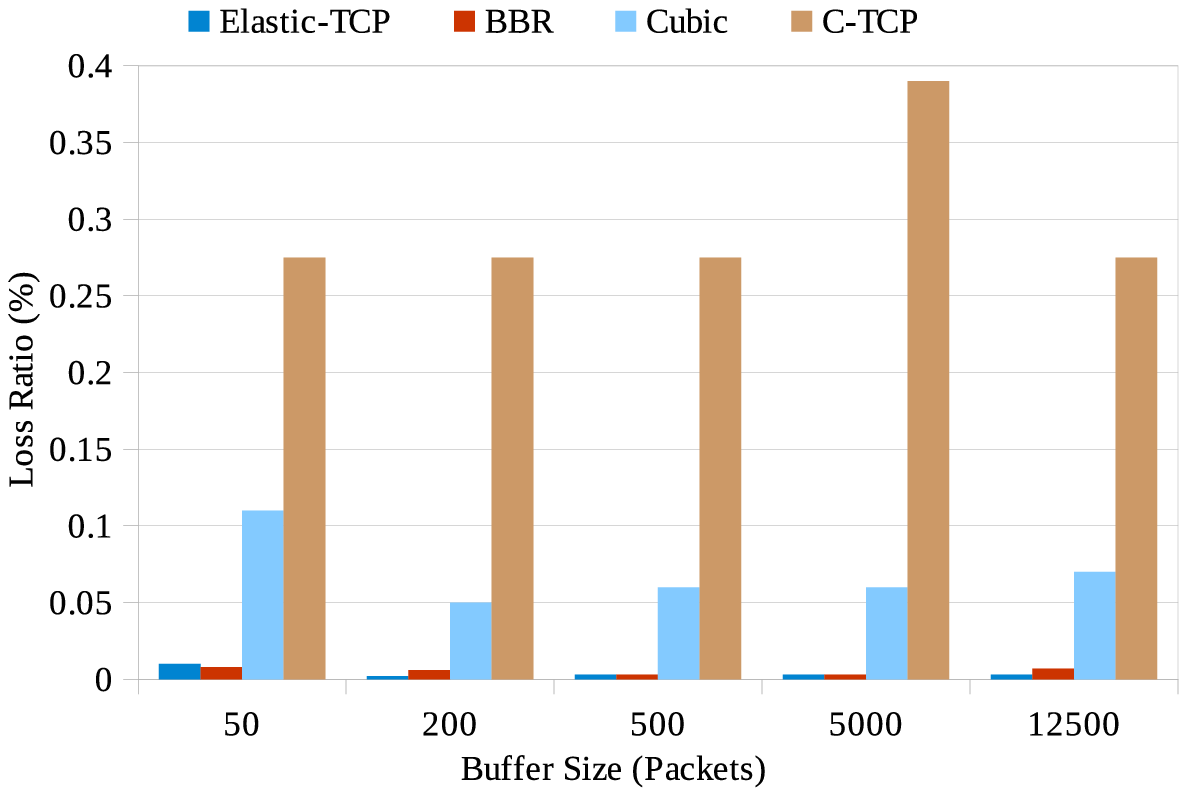}
			\label{fig:15}
		}
	\end{center}
	\caption{Single-flow scenario with different buffer sizes.}
	\label{fig:16}
\end{figure}

\begin{figure} [h!]
	\centering
	\begin{center}
		\hspace{2mm}
		\subfigure[Average throughput \& SD with CI 95\%.] 
		{
			\includegraphics[width=0.45\linewidth]{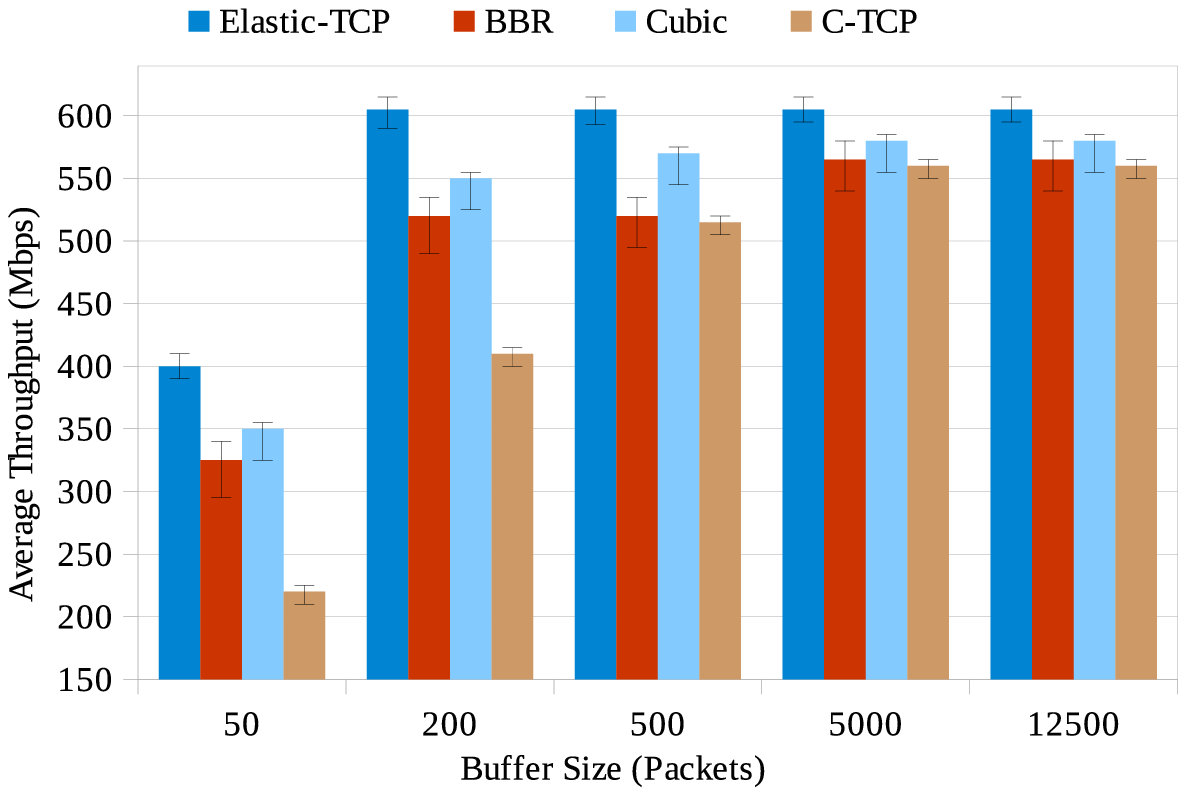}
			\label{fig:17}
		}\hspace{2mm}
		\subfigure[Loss ratio.] 
		{
			\includegraphics[width=0.45\linewidth]{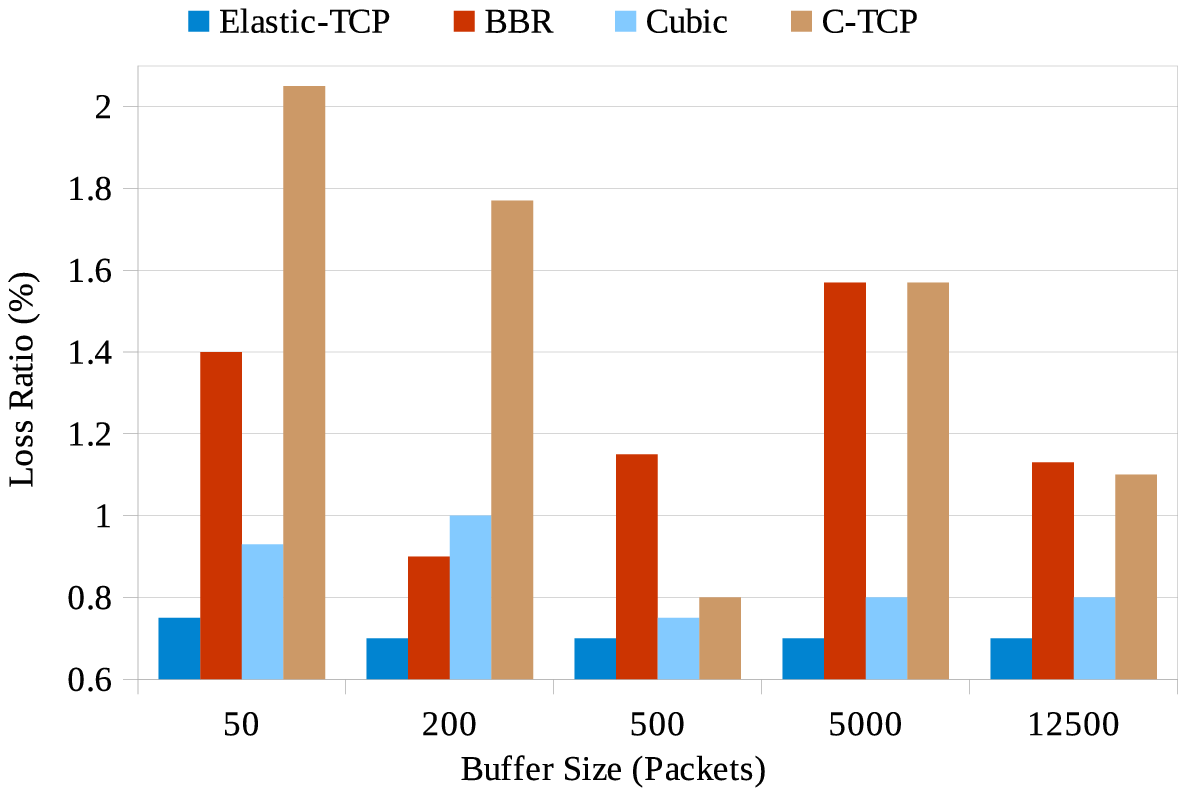}
			\label{fig:18}
		}
		\subfigure[Intra-fairness.] 
		{
			\includegraphics[width=0.45\linewidth]{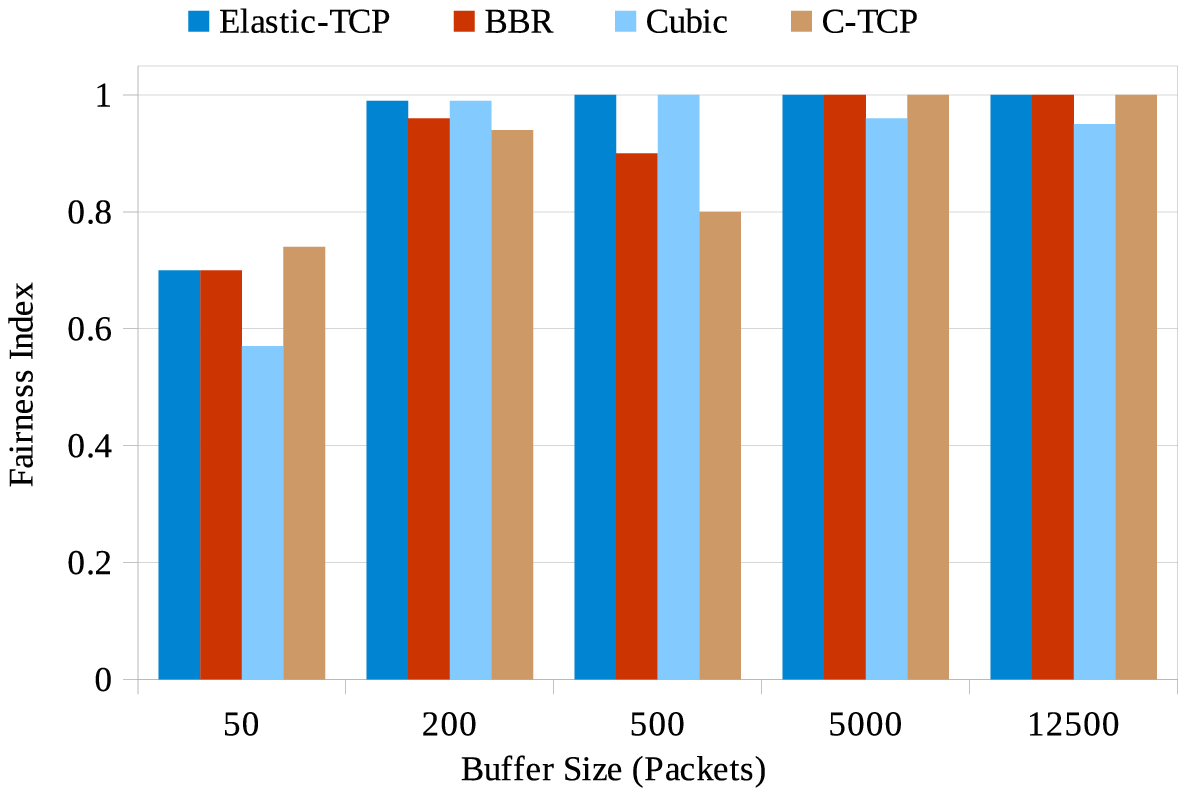}
			\label{fig:19}
		}
	\end{center}
	\caption{Multiple-flows scenario: four simultaneous FTP flows.}
	\label{fig:20}
\end{figure}

In the single-flow scenario, Elastic-TCP and TCP-BBR lose about 1 packet from every 10,000 packets (0.01\%), Cubic loses about 10 packet from every 10,000 packets (0.1\%), and C-TCP loses about 30 packets from every 10,000 packets (0.3\%), as shown in Figure \ref{fig:15}. In the multiple-flows scenario, in the cases of small buffers, Elastic-TCP and Cubic show the lowest loss ratio, where Elastic-TCP loses about 7 packets from every 1000 packets (0.7\%) and the Cubic loses about 10 from every 1000 packets (1\%) while TCP-BBR and C-TCP losses up to 1.4\% and 2.1\%, respectively. As for the large buffer scenarios, the loss ratio of all algorithms is between 0.8\% to 1.5\%, where the lowest loss ratio is provided by Elastic-TCP, as shown in Figure \ref{fig:18}.

\subsubsection{The fairness}
The examined TCP CCAs have achieved similar intra-fairness in most cases. In the case of 50 packets buffer, C-TCP seems fairer than the compared CCAs followed by TCP-BBR, Elastic-TCP, and Cubic. However, the difference between the higher fairness and the lower fairness measurment ranges from 2\% to 10\%, which is slightly acceptable.

\section{Summary and discussion of results} \label{disc} The results reveal that Elastic-TCP is able to achieve higher bandwidth utilization compared to other TCP CCAs, while it minimizes the loss ratio and maintains the fairness. Due to its unique function, the proposed Elastic-TCP shows less sensitivity to the changes of PER and buffer size. In general, it shows better performance compared to other TCP CCAs, which is considered a significant improvement in terms of bandwidth utilization. 

Based on simulation, Elastic-TCP improves: (1) up to 22\% in the case of single flow, (2) up to 40\% in the case of sequential multiple flows and (3) up to 50\% in the case of synchronous multiple flows. In the second scenario, which represents a real network case where the coexisting TCP flows are not synchronously established or terminated, Elastic-TCP utilizes up to 80\% of the available bandwidth while the others could not exceed 66\% in case of large buffer size. Moreover, Elastic-TCP achieves from 47\% to 66\% bandwidth utilization, in the case of small buffer size, while the bandwidth utilization of the compared TCP CCAs varies from 5\% to 29\%. With regards to the impact of synchronized losses among the competing flows, the third scenario is used to show the impact of these losses on the average throughput. Fortunately, Elastic-TCP improves the throughput up to 50\%, especially when the PER is high.

Furthermore, a testbed experiment is conducted to compare the performance of Elastic-TCP to the recent TCP CCAs available in the upcoming Linux kernel version 4.9, including Cubic, C-TCP, Agile-SD, and TCP-BBR. Indeed, TCP-BBR, which is recently developed by Google, is the most promising candidate to replace the current congestion control algorithms in the upcoming Linux kernel. However, the results show that the proposed Elastic-TCP can outperform Cubic, C-TCP, Agile-SD, and even TCP-BBR. Elastic-TCP improves the average throughput by up to 23\% over TCP-BBR, up to 14\% over Cubic and up to 81\% over C-TCP.

\section{Conclusion}
\label{Conc}
In this work, a novel RTT-independent and delay-based TCP CCA, namely Elastic-TCP, is proposed and evaluated. Elastic-TCP mainly contributes a new Window-correlated Weighting Function (WWF). Basically, the necessity of Elastic-TCP has been arisen by the inability of the existing CCAs in achieving full bandwidth utilization over high-BDP networks, especially when the used buffer is small and/or the packet losses are common. Further, a new Elastic-TCP module is designed, developed and attached to the NS-2 as a Linux-TCP module, which is ready for implementation into Linux kernel.  Thereafter, simulation and testbed experiments are carried out to examine the performance of Elastic-TCP compared to TCP-BBR, Cubic, C-TCP, and Agile-SD. Elastic-TCP introduces significant improvement in terms of bandwidth utilization especially over congested networks, where the available buffer at the bottleneck is small and the loss ratio is very high.

The utility of Elastic-TCP is maximized if the sender-side end systems are Linux-based, which is very likely since a large number of Internet servers are Linux-based. However, since Elastic-TCP is an algorithm, it is not bound to a specific operating system and it can be implemented in any operating system such as Windows, Macintosh, and Sun Solaris. 

Finally, the Elastic-TCP should be evaluated over satellite networks in order to take into account any potential issues. Also, there is a strong intention to examine the Elastic-TCP over wireless and mobile networks to study the impact of route changing and hand-off.

\section*{Acknowledgment}
This work is supported by Universiti Putra Malaysia and Al~Asmarya Islamic University - Libya.
\ifCLASSOPTIONcaptionsoff
  \newpage
\fi


\begin{IEEEbiography}[{\includegraphics[width=1in,height=1.2in,clip]{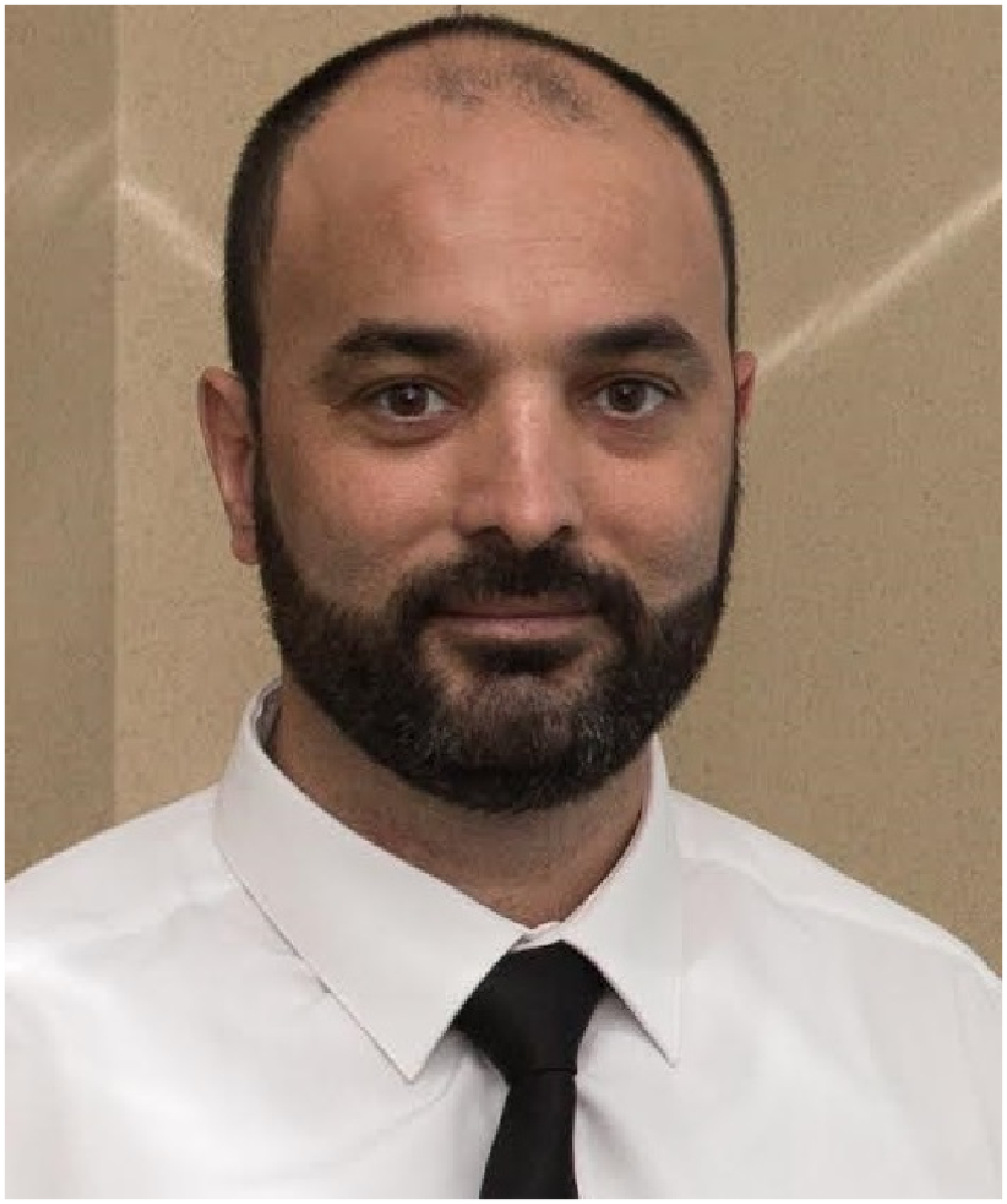}}]{Mohamed A. Alrshah}
(M'13--SM'17) received his BSc degree in Computer Science from Naser University - Libya, in June 2000. Then, he received his MSc and Ph.D degrees in communication technology and networks from Universiti Putra Malaysia (UPM) in May 2009 and Feb 2017, respectively. Now, he is an Assistant Professor (Senior Lecturer) in the Department of Communication Technology and Networks, Faculty of Computer Science and Information Technology, Universiti Putra Malaysia (UPM). He has published a number of articles in high-impact factor scientific journals. His research interests are in the field of high-speed TCP protocols, high-speed wired and wireless network, WSN, MANET, VANET, parallel and distributed algorithms, IoT and cloud computing.
\end{IEEEbiography}

\begin{IEEEbiography}[{\includegraphics[width=1in,height=1.2in,clip]{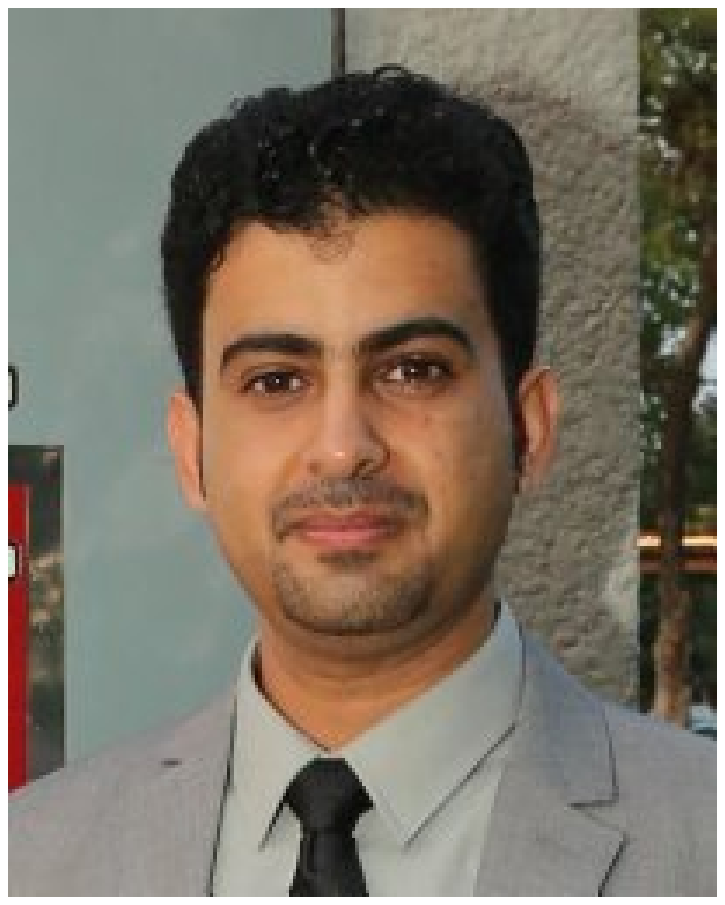}}]{Mohamed A. Al-Moqri}
received his BSc degree in Computer Science from Almustanseriah University - Iraq, in 2004. Then, he received his MSc and Ph.D degrees in communication technology and networks from Universiti Putra Malaysia in 2009 and 2016, respectively. Now, he is a Lecturer and Head of department of Information Technology in the Faculty of Computer Science and Information Technology, Azal University for Human Development, Sana'a, Yemen. He has published a number of articles in high-impact factor scientific journals. His research interests are in the field of high-speed TCP protocols, high-speed network, QoS, scheduling algorithms, admission control and wireless networks.
\end{IEEEbiography}

\begin{IEEEbiography}[{\includegraphics[width=1in,height=1.2in,clip]{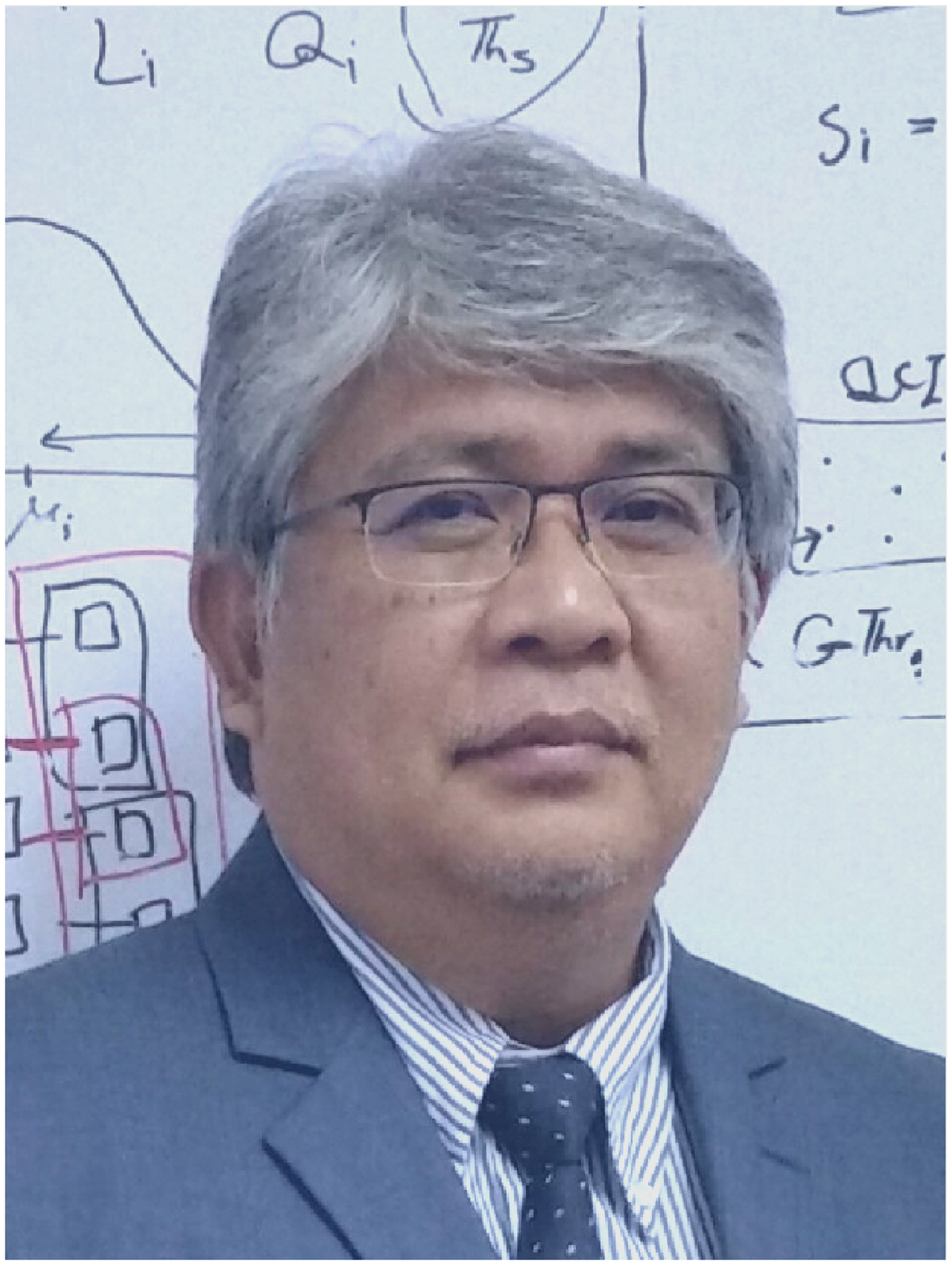}}]{Mohamed Othman}
(M'06--SM'18) received his Ph.D from the Universiti Kebangsaan Malaysia (UKM) with distinction (Best Ph.D Thesis in 2000 awarded by Sime Darby Malaysia and Malaysian Mathematical Science Society). Now, he is a Professor in the Department of Communication Technology and Networks, Faculty of Computer Science and Information Technology, Universiti Putra Malaysia (UPM). He is also an associate researcher at the Lab of Computational Science and Mathematical Physics, Institute of Mathematical Research (INSPEM), UPM. He published more than 160 International journals and 230 proceeding papers. His main research interests are in the fields of high-speed network, parallel and distributed algorithms, software defined networking, network design and management, wireless network (MPDU- and MSDU-Frame aggregation, MAC layer, resource management, and traffic monitoring) and scientific telegraph equation and modelling.
\end{IEEEbiography}
\vfill
\end{document}